\documentclass[pdfatex,sn-basic]{sn-jnl} 

\usepackage{booktabs}

\usepackage{subcaption}
\usepackage{graphicx}
\usepackage{array}

\usepackage{times}
\usepackage{latexsym}
\usepackage{listings}
\usepackage[T1]{fontenc}
\usepackage[utf8]{inputenc}
\usepackage{amsmath}

\usepackage{microtype}

\usepackage{inconsolata}

\usepackage{graphicx}
 \usepackage{multirow}

\usepackage{todonotes}

\usepackage{algorithm} 
\usepackage{algpseudocode} 
\usepackage{adjustbox}
\usepackage{geometry}
\usepackage{comment}
\usepackage{array}
\usepackage{graphicx} 
\algdef{SE}[SUBALG]{Indent}{EndIndent}{}{\algorithmicend\ }%
\algtext*{Indent}
\algtext*{EndIndent}
\usepackage{soul}

\usepackage{pgfplots}

\usepgfplotslibrary{groupplots}
\pgfplotsset{compat=1.8}
\usepgfplotslibrary{statistics}
\usepackage{xcolor}

\lstdefinestyle{pythonstyle}{
    language=Python,
    basicstyle=\tiny\ttfamily,
    breaklines=true,
    showstringspaces=false,
    commentstyle=\color{gray},
    keywordstyle=\color{blue},
    stringstyle=\color{green},
    frame=none,  % Removed frame
    numbers=none,
    breakatwhitespace=false,
    breakindent=0pt,
    tabsize=2
}

\newcommand{\corpus}{MultiAIGCD}

\begin{document}

%\title{\corpus{}: A \textbf{C}omprehensive dataset for AI Generated Code Detection Covering Multiple \textbf{La}nguages, \textbf{M}odels, \textbf{P}rompts, and \textbf{S}cenarios }

\title{\corpus{}: A Comprehensive dataset for AI Generated Code Detection Covering Multiple Languages, Models, Prompts, and Scenarios }

\author[1]{\fnm{Basak} \sur{Demirok}}\email{demirok@etu.edu.tr}

\author[2]{\fnm{Mucahid} \sur{Kutlu}}\email{mucahidkutlu@qu.edu.qa}

\author[3]{\fnm{Selin} \sur{Mergen}}\email{mergen@etu.edu.tr}
\affil[1]{\orgdiv{Department of Computer Engineering}, \orgname{TOBB University of Economics and Technology}, \orgaddress{\city{Ankara}, \country{Türkiye}}}

\affil[2]{\orgdiv{Department of Computer Science and Engineering}, \orgname{Qatar University}, \orgaddress{\city{Doha}, \country{Qatar}}}

\affil[3]{\orgdiv{Department of Computer Engineering}, \orgname{TOBB University of Economics and Technology}, \orgaddress{\city{Ankara}, \country{Türkiye}}}

\maketitle

\begin{abstract}

%With the rapid advancement of LLM models, they have become widely useful in various fields. While these AI systems can be used for code generation, significantly simplifying and accelerating the tasks of developers,  their use for students to do assignments has raised ethical questions in the field of education. In this context, determining the author of a particular code becomes important. In this study, we introduce \hl{AIGCodeSet},  a dataset for AI-generated code detection tasks, for the Python, Java, and Go programming languages. We obtain the problem descriptions and human-written codes from the CodeNet dataset. Using the problem descriptions, we generate AI-written codes with Llama-3.3-70B-Instruct-Turbo, Qwen2.5-Coder-32B-Instruct, GPT-4o, OpenAI o3-mini, Claude 3.5 Sonnet v2, DeepSeek-V3, and DeepSeek R1 models in three approaches: i) generating code from the problem description alone, ii)  generating code using the description along with human-written source code containing runtime errors,  and iii) generating code using the problem description and human-written code that resulted in wrong answers. Lastly, we conducted a post-processing step to eliminate  LLM output irrelevant to code snippets. Overall, \hl{AIGCodeSet} consists of 122,888 AI-generated and 32,190 human-written code snippets. We share our dataset with the research community to support studies on this important topic and provide performance results for baseline AI-generated code detection methods.

As large language models (LLMs) rapidly advance, their role in code generation has expanded significantly. While this offers streamlined development, it also creates concerns in areas like education and job interviews. Consequently, developing robust systems to detect AI-generated code is imperative to maintain academic integrity and ensure fairness in hiring processes.   
In this study, we introduce \corpus{}, a dataset for AI-generated code detection for Python, Java, and Go. From the CodeNet dataset's problem definitions and human-authored codes, we generate several code samples in Java, Python, and Go with six different LLMs and three different prompts. This generation process covered three key usage scenarios: 
(i) generating code from problem descriptions, (ii) fixing runtime errors in human-written code, and (iii) correcting incorrect outputs. Overall, \corpus{} consists of 121,271 AI-generated and 32,148 human-written code snippets. We also benchmark three state-of-the-art AI-generated code detection models and assess their performance in various test scenarios such as cross-model and cross-language. We share our dataset and codes to support research in this field. %\footnote{The URL for the dataset and code will be shared once the paper is accepted}. 
%We share this dataset to advance research in AI-generated code detection and provide baseline performance results.
\end{abstract}

% En son yazılacak

\section{Introduction}

With the rapid advancements in generative AI, large language models (LLMs) have become essential tools for coding. Studies show that AI-assisted development can significantly boost productivity, with developers reporting up to a 33\% increase when using AI-powered tools \cite{becker2023programming}. These tools also benefit students by offering solutions, explanations, and debugging assistance.  

However, AI-generated code raises concerns about academic integrity, plagiarism, security vulnerabilities, and potential skill degradation among developers \cite{ambati2024navigating, prather2023s}. Prior research studies report that LLMs may produce insecure, buggy, or inefficient code \cite{perry2023users, jesse2023large}, posing risks in real-world applications.  

Given these challenges, detecting AI-generated code has become a growing area of study \cite{hou2024large}. Existing research predominantly investigates LLMs' ability to generate code from problem definitions \cite{bukhari2023distinguishing, pan2024assessing}.  However, the  application of LLMs in debugging and error correction remains a critical, yet underexplored, area. Furthermore, the lack of a standardized dataset to assess model performance leads many researchers to create their own datasets for experiments, complicating systematic and fair comparisons across models. Consequently, there is a critical need for comprehensive benchmark datasets to advance research in this area.

In this study, we introduce \corpus{}, a
comprehensive dataset for AI generated code detection
covering multiple languages, models, prompts, and scenarios. In particular, \corpus{} covers  800 programming problems, three programming languages, six LLMs, three using scenarios, and three prompting approaches for code generation. We focus on Java, Python, and Go programming languages and identify 800 programming problems in IBM’s CodeNet \cite{puri2021codenet}. We collect up to five human-authored code samples from each category: correct submissions, submissions with runtime errors, and submissions that produce incorrect outputs. Next, using six different LLMs including \textit{Llama-3.3-70B-Instruct-Turbo, Qwen2.5-Coder-32B-Instruct, GPT-4o, OpenAI o3-mini, Claude 3.5 Sonnet v2, and DeepSeek-V3} and we generate codes in each language we target for the following  usage scenarios:  i) generating code from problem definitions,  ii) fixing code with runtime errors, and iii)  correcting incorrect outputs.  Furthermore, we employ three prompting strategies, namely \textit{Role} \cite{wang2023rolellm}, \textit{Lazy}, and \textit{Rephrase \& Respond} \cite{deng2023rephrase} to generate variants of each case. Lastly, we apply post-processing to remove irrelevant LLM outputs. \corpus{} ultimately comprises 32,148 human-written and 121,271 AI-generated code snippets.  

In our experiments, we benchmark three state-of-the-art AI-generated code detection models and evaluate their performance in various test environments including cross-model and cross-language. Our observations are as follows. 
OpenAI's ADA embeddings consistently yield the highest prediction accuracy across most scenarios. 
Furthermore, while the models demonstrated strong performance in detecting code generated from problem definitions, their accuracy significantly decreases when identifying AI-fixed code samples. Moreover, their performance also varies across programming languages. Lastly, while cross-model scenarios do not significantly impair detection accuracy, we observe a substantial decline  in cross-language setups.

%if the language and model of the training and test data match. However, their performance significantly decreases in cross-model and cross-language setups. 

The main contributions of our work are as follows.
\begin{itemize}
    \item We introduce \corpus{}, which covers 800 programming problems, three programming languages, six LLMs, three using scenarios, and three prompting approaches. \corpus{} is one of the few datasets for detecting code generated by artificial intelligence, including reasoning models such as OpenAI o3-mini. 
    \item We provide baseline results for three state-of-the-art models and assess their performance in several test scenarios including cross-language and cross-model.
    \item We share our code and data to support research in this important area\footnote{The URL for the data and code will be shared when the paper is published.}.
\end{itemize}

The rest of the paper is structured as follows: Section \ref{sec_rel} discusses related work. We explain details of dataset construction in Section \ref{sec_data}. In Section \ref{codingdiff}, we examine coding differences between LLMs and humans. We present experiments in Section \ref{sec_exp}. We provide limitations and ethical considerations in Section \ref{sec_limitations} and Section \ref{sec_ethic}, respectively.  We conclude in Section \ref{sec_conc}.

\section{Related Work}\label{sec_rel}

Despite relatively recent advances in LLMs for code generation, several  researchers worked on code generation from various aspects such as developing methods to detect AI-generated code \cite{xu2024detecting, oedingen2024chatgpt,bukhari2023distinguishing, 10.1145/3626252.3630826, idialu2024whodunit}, evaluating existing AI-generated code detectors \cite{pan2024assessing, wang2023evaluating}, and detecting vulnerabilities in AI-generated codes \cite{cotroneo2025devaic,wang2024your}. The studies differ in terms of the languages investigated, the number of problems,  size of datasets, LLMs employed, and source of the data.

 %The comparison of research findings is complicated due to the limited dataset resources comparing the diversity in LLMs, programming languages, and coding problems. 
 There are few studies introducing datasets specifically for AI-generated code. \citet{tihanyi2023formai} created a dataset containing 112,000 C language code samples generated using GPT-3.5 Turbo. Similarly, \citet{wang2024your} developed the CodeSecEval dataset, utilizing several LLMs, including multiple GPT models, CodeLlama-7B, and Claude 3 Opus, for Python code. Both datasets primarily address the security of generated code rather than the detection of AI-generated code.

Research on AI-generated code has mainly focused on specific programming languages. Python has been the most studied \cite{pan2024assessing, cotroneo2025devaic, oedingen2024chatgpt, idialu2024whodunit, wang2024your}, followed by C \cite{bukhari2023distinguishing, tihanyi2023formai}, and Java \cite{10.1145/3626252.3630826}. Some studies have explored multiple languages; for instance, \citet{xu2024detecting} examined C, C++, C\#, Java, JavaScript, and Python, while \citet{wang2023evaluating} investigated a set including Ruby, JavaScript, Go, Python, Java, and PHP. In our study, we  focus on Python, Java, and Go due to their popularity.

In creating a dataset for AI generated code detection, it is important to determine the problems to be covered and how to obtain the human-written codes. Existing studies use various resources for problem definitions and human-written codes. For instance,  \citet{pan2024assessing} utilize codes and problems from Kaggle, Quescol, and LeetCode; \citet{10.1145/3626252.3630826} incorporate CodeWorkout; \citet{xu2024distinguishing} operate CodeSearchNet; and \citet{idialu2024whodunit} use CodeChef.
Similar to our approach, \citet{xu2024detecting} utilize the CodeNet dataset, applying criteria such as selecting code with line lengths between 10 and 100, an alphanumeric character fraction greater than 0.25, and excluding all comments and duplicate files. In our study, we focus on problems which has sufficient successful and unsuccessful (i.e., runtime error or incorrect output) codes.

Regarding the LLMs used for AI-generated code studies, OpenAI's models are the most popular ones such as ChatGPT \cite{wang2023evaluating, liu2024refining, suh2024empirical,pan2024assessing, 10.1145/3626252.3630826}, Davinci  \cite{xu2024detecting, bukhari2023distinguishing}, GPT-4 \cite{idialu2024whodunit, suh2024empirical},  and tools incorporating GPT-4, such as GitHub Co-Pilot and Microsoft Co-Pilot \cite{cotroneo2025devaic}. Other LLMs include  Google's Bard \cite{ambati2024navigating}, Gemini \cite{suh2024empirical, cotroneo2025devaic}, CodeLlama-7B and Claude 3 Opus \cite{wang2024your}. 

While the AIGCodeSet \cite{demirok2024aigcodeset}  is the closest in scope, covering runtime error fixes and output corrections, it is restricted to Python and contains only 2,282 generated codes from CodeLlama, Codestral, and Gemini 1.5 Flash. Our research, however, significantly expands upon this by incorporating a substantially larger number of code samples, alongside a wider array of programming languages, a broader selection of LLMs, and a greater diversity of prompts.

\textbf{Table \ref{tab_related_work}} provides a comparison of notable datasets in the literature for AI-generated code detection.  
Our study differs from existing research in two key aspects. First, to the best of our knowledge, there is only one dataset \cite{guo2025codemirage} that covers reasoning models, while we provide code samples for OpenAI o3-mini.   Second, while the majority of prior research has focused on LLMs generating code from problem definitions, there has been limited investigation into AI-generated code within scenarios where LLMs are utilized to fix errors in human-written code for specific programming problems.  Overall, \corpus{} is the most comprehensive dataset for the AI-generated code detection problem, covering six LLMs, three programming languages, three prompts, and three usage scenarios.

\begin{table*}[!htb]
\footnotesize
\label{Tab_prompts}
  \begin{tabular}{ | p{3cm}  | p{3cm} | p{3cm} |p{2.2cm} |l |l |}
    \hline
    & LLMs &  Languages & Tasks & \# H & \# LLM  \\ \hline
\cite{wang2023evaluating}   & ChatGPT & Ruby, Javascript,
Go, Python, Java and PHP & Generation & 226,500 & 226,500   \\ \hline

\cite{yang2023zero}   & text-davinci-003, GPT-3.5, and GPT-4 & Python and Java & Generation & 237 & 711*   \\ \hline

\cite{xu2024detecting} & OpenAI’s text-davinci-003 & C, C++, C\#, Java, JavaScript, and Python & Generation\& Code Translation & 5,214 & 5,214   \\ \hline
\cite{bukhari2023distinguishing} &  code-cushman-001, code-davinci-001 and code-davinci-002 &  C & Generation & 28 & 30  \\ \hline
\cite{pan2024assessing} &  ChatGPT & Python & Generation & 5,069 & 65,897 %5.069*13(different prompts)  
\\ \hline
\cite{idialu2024whodunit} &  GPT-4 & Python & Generation & 798 & 798  \\ \hline
\cite{oedingen2024chatgpt} &  ChatGPT & Python & Generation & 15,700 & 15,700  \\ \hline
\cite{rahman2024automatic} &  Claude 3 Haiku & Python & Generation & 33,199* & 33,199* \\ \hline
\cite{xu2024distinguishing} &  ChatGPT & Python and Java & Generation & 612,000 & 500,000  \\ \hline
\cite{pham2024magecode} &   GPT-4-turbo, Gemini-pro-1.0, and Code-bison-32k & Python & Generation & 81,000 & 45,000 \\ \hline
\cite{suh2024empirical} &  Gemini Pro, Starcoder2-
Instruct (15B) , GPT-4, ChatGPT & C++, Python and Java & Generation & * & 29,591*  \\ \hline
\cite{ye2024uncovering} &  CodeLlama, StarChat, GPT-3.5 and GPT-4 & Python & Code Rewriting & 3,209 & 3,209  \\ \hline
\cite{bulla2024ex} &  ChatGPT & Java & Generation & 518 & 518  \\ \hline
\cite{gurioli2024you} &  StarCoder2 & C++, C, C\#, Go, Java, JavaScript, Kotlin, Python, Ruby, and Rust & Generation and \& Code Translation & 60,624  & 60,623  \\ \hline
\cite{xu2025codevision} &  GPT-3.5, GPT-4 & C, C++, Go, Java, Python, and Ruby & Generation & 12,709 & 15,633  \\ \hline
\cite{demirok2024aigcodeset} &  Codestral, Codellama, and Gemini Flash 1.5 & Python & Code Generation, Runtime Error Fix, and Correcting Output &  4,755 & 2,828   \\ \hline

\cite{gunawardhana2025approach} &  GPT-4o & Firmware code for Arduino platform & Generation &  90 & 90   \\ \hline

\cite{orel2025codet} &  GPT-4o, CodeLlama (7B), Llama3.1 (8B), CodeQwen 1.5 (7B), ve Nxcode-orpo  & C++, Java, and Python & Generation &  252,886	& 246,581   \\ \hline

\cite{guo2025codemirage} & GPT-4o-mini, o3-mini, Claude3.5-Haiku, Gemini-2.0-Flash, Gemini-2.0-Flash-Thinking- Experimental, Gemini-2.0-
Pro-Experimental, DeepSeek-V3, DeepSeek-R1, Llama-3.3-70B, and Qwen-2.5-
Coder-32B & C, C++, C\#, Go, HTML, Java, JavaScript, PHP, Python, and Ruby & Generation \& Paraphrasing & 10,000 & 200,000 \\ \hline

\cite{pordanesh2025hiding} &  GPT-4o, Gemini 1.5 Flash, and Claude 3.5 Sonnet &  Python & Generation &  6,026	& 6,026   \\ \hline

\cite{orel2025textttdroidresourcesuiteaigenerated} &Llama, CodeLlama, GPT-4o, Qwen, IBM Granite, Yi, DeepSeek, Phi, Gemma, Mistral, Starcoder (A total of 43 language models with variations) & C++,C, C\#, Go, Java, JavaScript, and Python  & Generation \& Adversarial \& Code Edit & * & * \\ \hline

\corpus{} & Qwen 2.5 Coder, Llama 3.3, DeepSeek V3, Claude 3.5 Sonnet v2, GPT-4o, OpenAI o3-mini & Java, Go, and Python & Code Generation, Runtime Error Fix, and Correcting Output & 32,148 &  121,271   \\ \hline
  \end{tabular}
  \caption{Comparison of datasets in the literature for detecting AI-generated code.  \textit{\#H} denotes the number of human-authored code samples, while \textit{\#LLM} represents the number of LLM-generated code samples. \textit{Cells containing * indicate information not explicitly stated in the original work.}}
  \label{tab_related_work}
\end{table*}

%Lastly, although some studies employ multiple prompts for different objectives, we utilize three distinct prompt types, consistently generating code for the same purpose, as specific instructions can influence the generated results.

%Our study diverges from existing research in two significant aspects. First, to the best of our knowledge, no prior work has utilized Codestral or Code Llama 34B for detecting AI-generated code. Second, while most previous studies have focused on generating code from problem definitions, none have explored detecting AI-generated code specifically in scenarios where LLMs are employed to correct errors in provided human-written code for a programming problem.

%Other models frequently employed are OpenAI's Davinci models \cite{xu2024detecting, bukhari2023distinguishing}, GPT-4 \cite{idialu2024whodunit, suh2024empirical}, and  Additionally, Google's Bard (an earlier version of Gemini) \cite{ambati2024navigating} and Gemini itself \cite{suh2024empirical, cotroneo2025devaic} are also have taken into consideration.

\section{\corpus{}}\label{sec_data}
In developing a dataset for detecting AI-generated code, the main requirement is to include both human-authored and AI-generated code samples. However, to ensure the dataset is of high quality and supports effective model training and reliable evaluation, several key considerations must be addressed. Specifically, it is essential to cover a wide range of coding problems and styles. Moreover, including both human-authored and AI-generated code for the same problem will allow for a more effective comparison of their differences. In addition, it is important to consider the level of involvement of AI tools and how they are used, as different individuals may use these tools in varying ways. Lastly, the dataset should cover various LLMs, as individuals may rely on different tools to solve their coding problems. Now, we explain our approach to constructing \corpus{} that aims to meet these objectives.

\subsection{Acquiring Human Written Codes}

Following prior work \cite{xu2024detecting}, we utilize IBM’s CodeNet dataset \cite{puri2021codenet} as a source for human-written code. %\footnote{It has CDLA Permissive v2.0 license which allows us to share the data}. 
Notably, these submissions date back to 2021, predating the emergence of most widely used code-assistant LLMs.

 This dataset contains 14 million code examples spanning approximately 4,000 coding problems across 55 programming languages. For each problem, there exists multiple submissions, allowing us to capture a range of coding styles. Each submission is assigned a status, which can be: (i) accepted, (ii) compile-time error, (iii) runtime error, (iv) wrong answer, or (v) time limit exceeded.  However, to align with our objectives and generate corresponding AI-generated versions at a reasonable cost, we selectively filter the dataset, retaining only the code samples most relevant to our study.

The official webpage of the CodeNet dataset\footnote{\url{https://developer.ibm.com/exchanges/data/all/project-codenet/}} provides a subset of the data called the “Python Benchmark” which includes 800 coding problems. %This benchmark dataset contains only accepted Python solutions along with their corresponding problem descriptions.  
We selected these 800 problem descriptions and, for each problem, extracted up to five submissions from the corpus across the accepted, runtime error, and wrong answer statuses in Python, Java, and Go. Consequently, we have a maximum of 15 submissions per problem (5 submissions x 3 statuses). However, the number of sampled submissions  may be fewer for some problems due to the limited number of submissions. Using submissions with various statuses enables us to include different coding styles and solutions, including incorrect ones. We leave other submission types, e.g., time limit exceed, as future work.

Overall, we sampled a total of 33,286 (11,873 Java -  12,000 Python - 9,413 Go) code snippets from the CodeNet dataset, covering various coding problems and a wide range of correct and incorrect solutions written by different individuals.

\subsection{Creating AI-Generated Code Dataset}

To obtain AI-generated code samples, we employ six LLMs: i) Llama-3.3-70B-Instruct-Turbo\footnote{\url{https://huggingface.co/meta-Llama/Llama-3.3-70B-Instruct}}, ii) Qwen2.5-Coder-32B-Instruct\footnote{\url{https://huggingface.co/Qwen/Qwen2.5-Coder-32B-Instruct}}, iii) GPT-4o, iv) DeepSeek-V3\cite{liu2024deepseek}, v) o3-mini, and vi) Claude 3.5 Sonnet v2\footnote{\url{https://www.anthropic.com/claude/sonnet}}. We accessed these models through various API platforms: Llama, Qwen, and DeepSeek-V3 were called via Together API\footnote{\url{https://together.ai/}}, Claude 3.5 Sonnet through Anthropic's batch API\footnote{\url{https://www.anthropic.com/api}}, GPT-4o and o3-mini through OpenAI's batch API\footnote{\url{https://platform.openai.com/}}.

For each coding problem and LLM, we generate code for three usage scenarios: 
\begin{itemize}
    \item \textit{Scenario\textsubscript{Scratch}}: Generating code from scratch for a given problem 
    \item \textit{Scenario\textsubscript{Runtime}}:  Fixing human-written code that results in a runtime error 
    \item \textit{Scenario\textsubscript{Output}}:  Correcting human-written code with incorrect output. 
\end{itemize}

Furthermore, we repeat the process for all languages we focus on and use three different prompts for each case. In some cases, we were unable to generate code due to the absence of available human-authored examples or the inability of LLMs to produce code for specific cases. Nevertheless, we could generate maximum 162 (=3 programming languages x 3 code statuses x 6 LLMs x 3 prompts) code samples for each problem.  

%This approach allows us to cover a range of AI-driven code-generation scenarios.

Regarding the prompts, we use the problem descriptions from CodeNet to define each coding task.  
%As part of preprocessing, we replaced multiple instances of '\textbackslash n' in the original descriptions with single '\textbackslash n's to improve readability. 
When prompting the model to fix a given code's running time error or its incorrect output, we randomly select one of the code snippets with the relevant status and use it for all LLMs and programming languages. 
We first did a pilot study to design effective prompts and explored the studies on prompting \cite{schulhoff2024prompt}. For instance, we observed that LLMs tend to use triple backtrips and the name of the programming language at the very beginning of the generated codes. %, this may stems from some LLMs are optimized for structured outputs. 
As this will make their detection easy, we modified our prompts to prevent generating such an output.  We used three different prompting approaches to design our prompts:

\begin{itemize}
    \item \textbf{Lazy Prompting. } We use a simple prompt with minimal directions to tailor the output. 
    \item \textbf{Role Prompting. } We apply the role prompting technique \cite{wang2023rolellm} where a specific role, in our case an expert programmer, is assigned to the LLM.
    \item \textbf{Rephrase and Respond.} \citet{deng2023rephrase} propose asking LLMs to rephrase a given prompt before generating the output. We also apply the same method and ask models to rephrase our initial prompt. \textbf{Table \ref{tab_prompts_for_prompts}} in Appendix provides the  prompts used to generate Rephrase and Respond prompts for each scenario. Note that we do not permit LLMs to rephrase the problem descriptions or the human-authored codes, in order to preserve the integrity of the code generation objective. We use the generated prompts for code generation. %Since each LLM might generate a different prompt, we use different prompts for each LLM. 
\end{itemize}
    
\textbf{Table \ref{Tab_code_prompt}} in Appendix provides the  prompts used for each scenario. In total, we generated 124,434 
(43,110 for Java, 38,124  for Go, and 43,200 for Python) code snippets, covering six distinct LLMs, three usage scenarios, three prompts, and 800 coding problems. %In the DeepSeek R1 model, where we could not get results from all problems and prompt types due to technical issues we faced (e.g., the )
%high demand and slow code generation, we produced a total of 1,568 codes (317 for Java, 304  for Go, and 965 for Python).
% bunu diyoruz zaten yukarıda

\subsection{Post-processing}

After generating the codes using the models, we performed a quality control check to ensure their validity and identified the following issues in some of the outputs: (i) failure to produce any code, (ii) inclusion of code written in C-family languages, and (iii) presence of outputs that do not follow our prompts such as code explanations and triple backticks. Therefore, we took the following steps to ensure the quality of the generated codes. Firstly, we removed the triple backtick portion of the code. To ensure syntactic validity of Java codes, we used  JavaLang parser\footnote{\url{https://pypi.org/project/javalang/}}  to parse the codes. % and accepted the ones that can be parsed. %We did not apply any regex-based heuristics to recover unparsable but potentially valid fragments. 
Any code that failed parsing was  discarded. 
 We applied this filtering step to both human-written and LLM-generated code samples.

Regarding Python codes, we employed AST\footnote{\url{https://docs.python.org/3/library/ast.html}} parser to ensure syntactic validity, similar to our Java approach, discarding unparsable samples.
For Go code, due to the absence of a robust and easily integrable Go parser for our processing pipeline, we implemented a set of heuristic rules based on language-specific keywords to verify that the code snippets were indeed written in Go.

%we did not apply formal syntax validation due to the lack of a robust and readily available Go parser that could be integrated into our processing pipeline. Instead, we defined a set of heuristic rules based on the presence of language-specific keywords to ensure that the snippet was written in Go. 

After  filtering the problematic code snippets, we obtained 121,271 LLM-generated %(except DeepSeek R1, for DeepSeekR1 we obtained 1,577 codes) 
and 32,148 human code snippets in total. The final statistics for \corpus{} are shown in \textbf{Table \ref{Tab_stats}}. 
 
% Furthermore, although the prompts specifically requested only code as output, some responses included additional explanations, either as standalone text or embedded within the code as comments. To ensure the dataset accurately reflects the intended challenge of detecting AI-generated code, we manually removed any explanations provided above or below the code. However, comments embedded within the code blocks were retained, provided they were appropriately marked as comment lines.

%The author scanned each of the 2,852 generated code snippets. Some files were empty, contained C-family language code instead of Python, or included meaningless characters, sentences, numbers, or dots. These snippets were excluded from the dataset, resulting in a final set of 2,824 Python code snippets. Each generated code distraction info stated in metadata available from *link*. When checking the code snippets, we realize that some of the generated codes are not completed or have obvious incorrect indentation that will be resulted in runtime error. We include even this codes to dataset because we already know that they may produce incomplete or wrong codes, that's why we perform ai-generated code detection.

%\setlength\tabcolsep{1.pt}
\begin{table*}[!htb]
%\small
\centering
  \begin{tabular}{ | l | l | l | l |l |l |  l | l |l  | l |}
    \hline
  
  & & Human & Qwen &  Llama & DS-V3 & Claude & GPT-4o & o3-mini  & \textbf{Total}\\ \hline
  
  \multirow{3}{*}{\rotatebox{90}{Java}} & 
  
  GS/A & 3985 &  2380 & 2398 & 2369  & 2397 & 2354 & 2153 &  18,036   \\ 
  
  & FR/R & 3857 &  2322 & 2333 & 2329  & 2372 & 2378 & 2117 & 17,708   \\ 
  
  & CO/W & 3984 &  2361 & 2355 & 2364  & 2391 & 2076 & 2091 &  17,913   \\ \hline \hline

  \multirow{3}{*}{\rotatebox{90}{Python}} & 
  
  GS/A & 3943 &  2398 & 2400 & 2399  & 2398 & 2380 & 2154 &  18,072  \\ 
  
  & FR/R & 3078 &  2340 & 2381 & 2358  & 2397 & 2388 & 2167 & 17,109   \\ 
  
  & CO/W & 3888 &  2374 & 2398 & 2369  & 2399 & 2390 & 2145 & 17,963   \\ \hline \hline

  \multirow{3}{*}{\rotatebox{90}{Go}} &
  GS/A & 4000 &  2400 & 2400 & 2377  & 2400 & 2400 & 2021 & 17,998 \\ 
  
  & FR/R & 1755 &  1584 & 1578 & 1570  & 1587 & 1587 & 1509 &  11,170  \\ 
  
  & CO/W & 3658 &  2367 & 2367 & 2367  & 2367 & 2367 & 1957 & 17,450  \\ \hline \hline
  
  \multicolumn{2}{|c|}{\textbf{Total}} & 32,148 &  20,526 & 20,610 & 20,502  & 20,699 & 20,635 & 18,299 & \textbf{153,419}   \\ \hline
  
  \end{tabular}
  \caption{Data Distribution after data elimination of \corpus{}. GS/A: Generate from scratch for LLM, Accepted status for human; FR/R: Fix runtime error for LLM, Runtime error for human, CO/W: Correcting the output for LLM, Wrong answer for human,  DS: DeepSeek} 
  \label{Tab_stats}
\end{table*}

\section{Qualitative Analysis}\label{codingdiff}

\subsection{Code Samples}
In order to provide more insight into our dataset, we provide sample Python codes for average selection problem in \textbf{Table \ref{Tab_code}} in Appendix.  These samples represent each LLM we use, along with one human-authored code. We observe that each LLM generates a different code for the same problem in terms of variable and function naming, algorithm development,  the length of codes, and writing styles. For instance, while the human-authored code has multiple consecutive blank lines, perhaps to increase the readability, none of the LLM-generated codes have consecutive blank lines.
%Specifically, the problem is as follows.
%\textit{For an integer n not less than 0, let us define f(n) as follows: }\\
%$f(n) = 1 (if n < 2) $ \\
%$ f(n) = n f(n-2) (if n \geq 2)  $ \\
%\textit{Given is an integer N. Find the number of trailing zeros in the decimal notation of f(N).}

%\textbf{Table \ref{Tab_code}} provides code samples in our dataset for the problem for all LLMs we use for the three different usage scenarios. For each scenario, we also provide an example human written code with a the corresponding status. 
 
\subsection{Coding Style Differences}
\begin{figure*}[!htb]
\centering
\includegraphics[width=0.45
\linewidth]{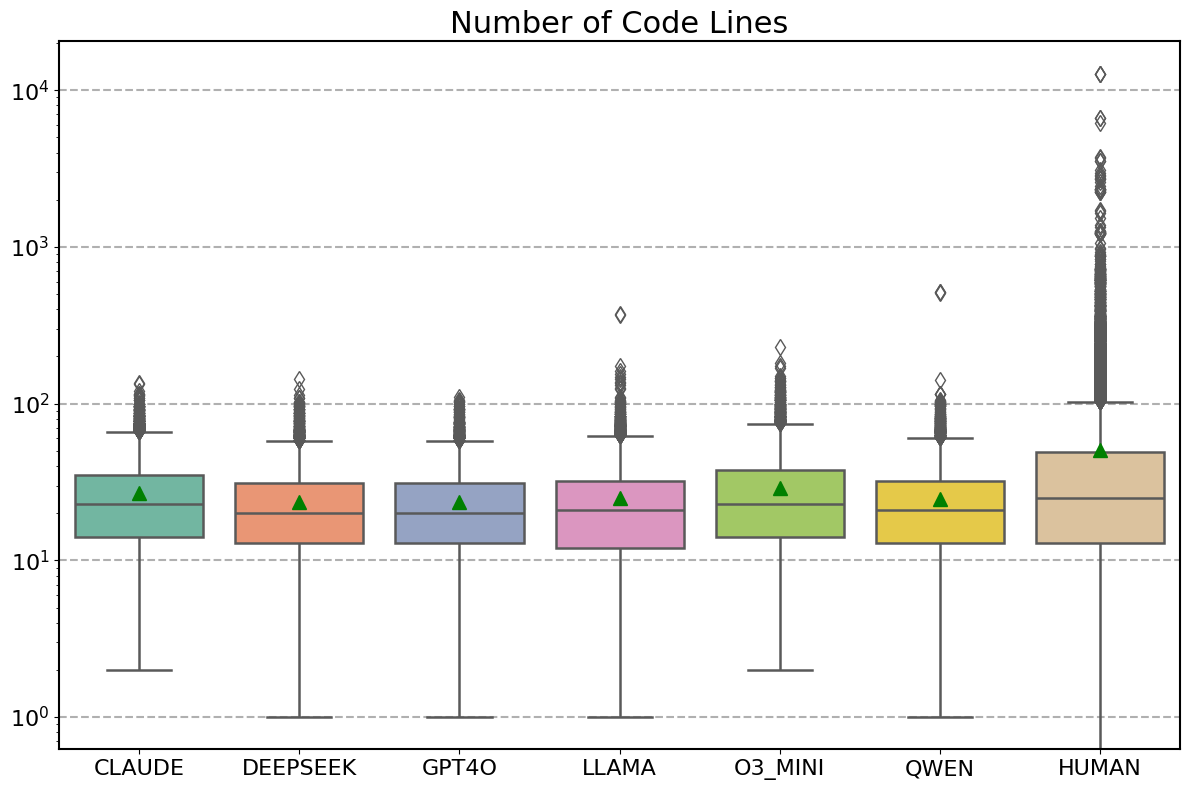}
\includegraphics[width=0.45\linewidth]{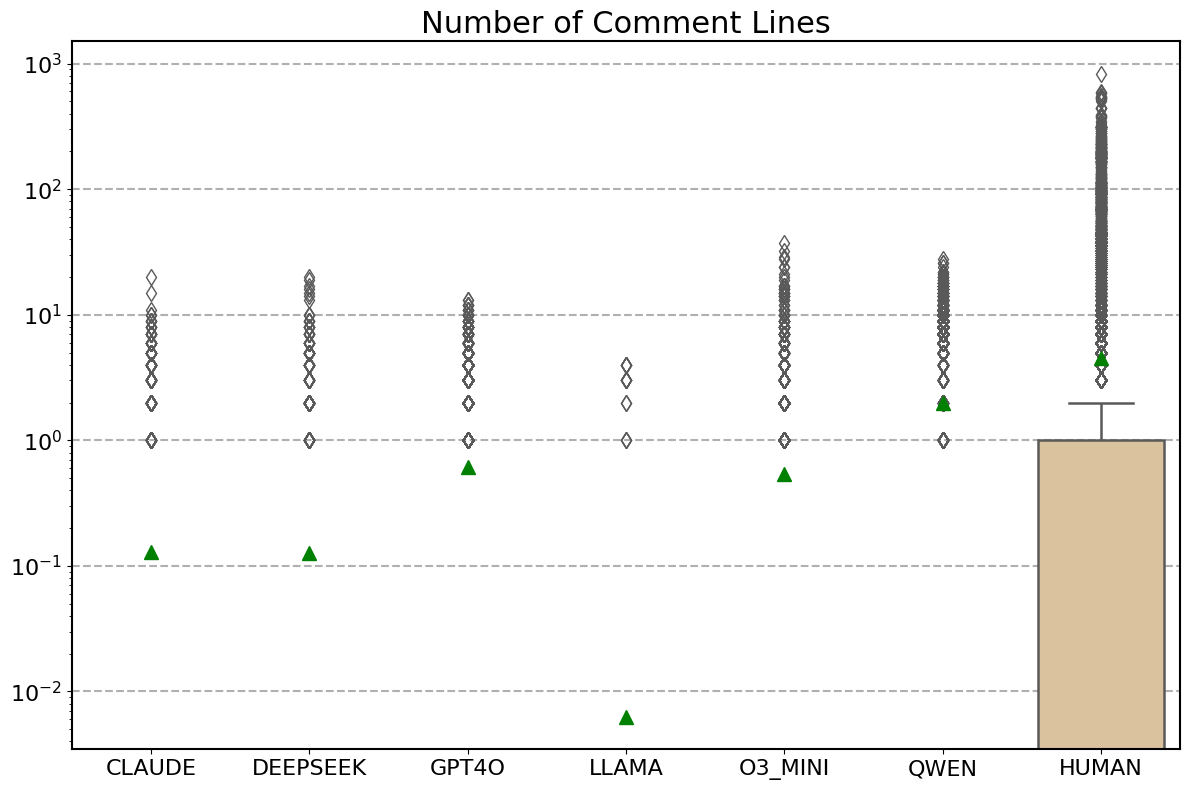}
\includegraphics[width=0.45\linewidth]{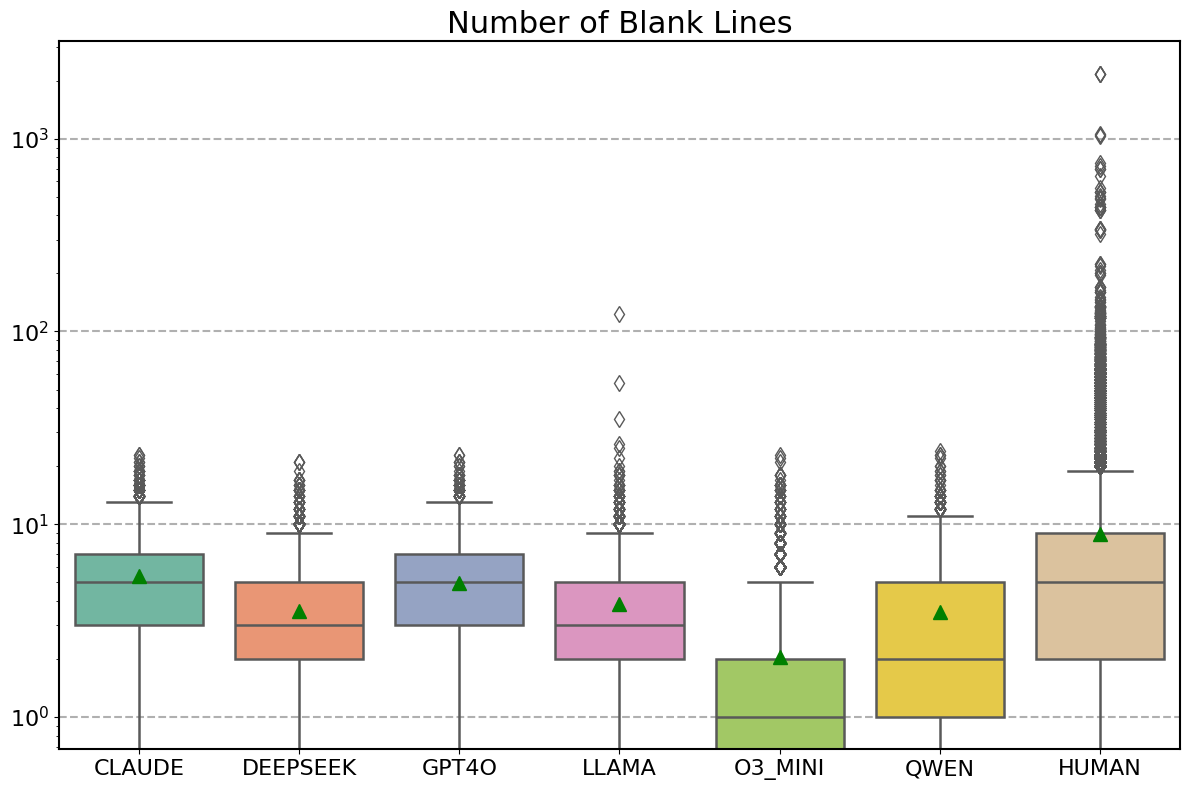}
\includegraphics[width=0.45\linewidth]{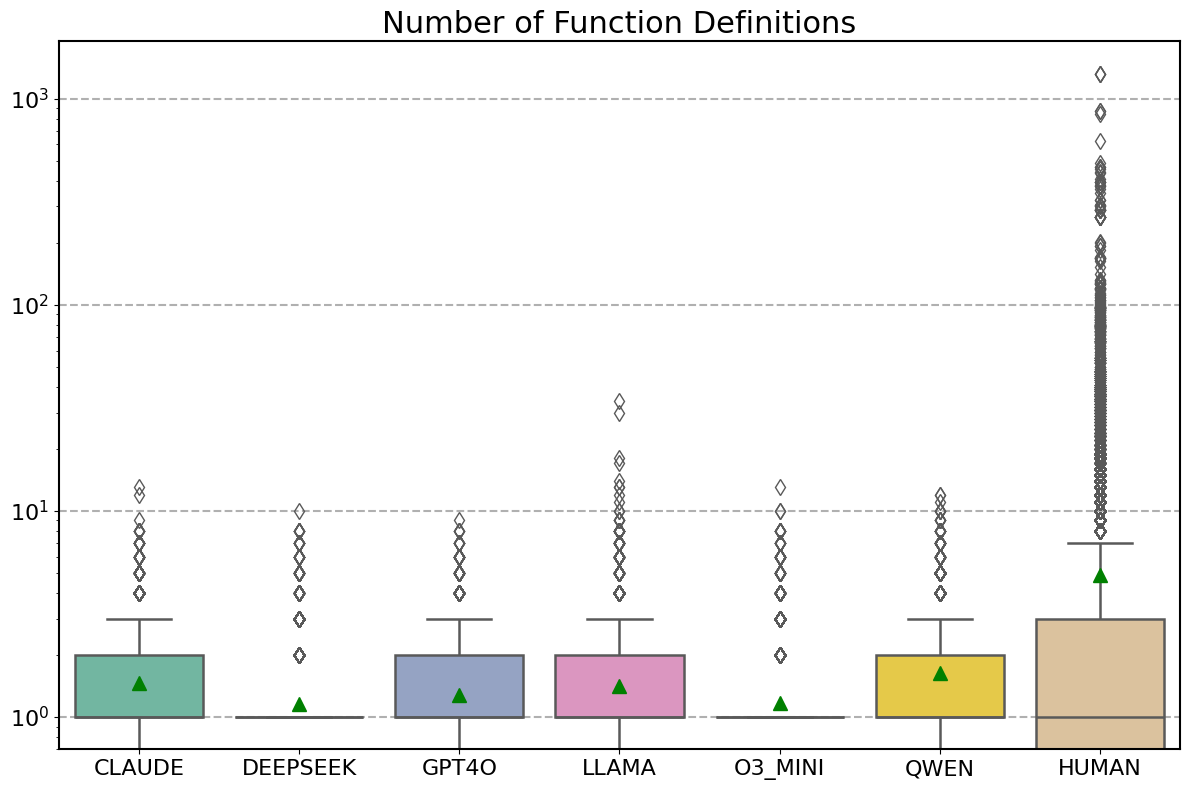}
\caption{The comparison of human-authored and LLM-generated codes in \corpus{} from various aspects including i) the total number of lines, ii) the number of comment lines, iii) the number of blank lines, and iv) the number of function definitions. The green marker represents the average in each distribution. The y-axis is shown in log scale for better visualization.}
\label{fig:llm}
\end{figure*}

To examine the general differences between AI-generated and human-authored code, we calculate the average number of lines, blank lines, comments, and function definitions for human-authored and AI-generated codes separately. For AI-generated samples, we focus solely on codes generated from scratch to better capture distinctions in LLM-generated codes. 
The results are presented in \textbf{ Figure \ref{fig:llm}}.

We observe that human-written codes tend to be longer compared to those generated by LLMs. %However, this might be because human-authored codes have also higher number of comments and blank lines on average. \todo{sanırım bunlar göz ardı edilmişti number of code lines'da? öyle ise bu yorum geçersiz.}
In addition, human-authored codes have more outliers in all cases, suggesting that humans are more creative than LLMs in writing codes. Regarding the presence of comments, the median of the number of comment lines in human-authored codes is just one but there are outliers that have extreme amount of comments. In most LLM-generated code snippets, comments are absent, as we explicitly instructed the models to omit them. 

Nonetheless, some outputs still include comments despite these directives. Regarding function definitions, LLMs typically define at least one function, whereas several human-authored snippets contain none. However, human-written code is more likely to include multiple functions. Finally, we observe notable differences in coding style across LLMs. For instance, O3-mini tends to include fewer blank lines on average, and both DeepSeek and O3-mini rarely define multiple functions compared to other models. 
%Among the LLMs, DeepSeek R1 notably generates more compact codes with fewer comments, function definitions, and blank lines.

\subsection{Code Accuracy}
In this section, we investigate the functional correctness of code generated by LLMs to analyze their behavioral patterns beyond stylistic or semantic characteristics. To capture the complete spectrum of model behaviors, our analysis includes all generated outputs without the application of the previously described filtering steps. 

%
%\begin{figure}[htb!]
%\centering
%\includegraphics[width=0.8
%\linewidth]{figures/histogram_io.png}
%\caption{histogram}
%\label{iohist}
%\end{figure}

To assess the functional correctness of the generated code, we executed each solution against predefined test cases derived from the input-output examples provided in the problem descriptions. Notably, 8 of the 800 problems lack these examples. % and were consequently excluded from this execution-based analysis.
The execution outcomes are classified into the following six categories:
\begin{itemize}
    \item \textbf{Pass.} The code successfully processes all test cases and produces the correct output.

    \item   \textbf{Compile Error.} The code fails during compile time (valid only for Java and Go).

   \item  \textbf{Incorrect Output.} The code completes execution without error but yields a result that does not match the expected output.
    
   \item \textbf{ Runtime Error.} The code fails to complete execution due to an error.

\item \textbf{Timeout. }The program exceeds the five-second execution time limit.

\item  \textbf{Missing.} The response contains no executable code or there is no input-output examples, as previously mentioned.  

\end{itemize}

\begin{figure*}[!htb]
\centering
\includegraphics[width=0.75
\linewidth]{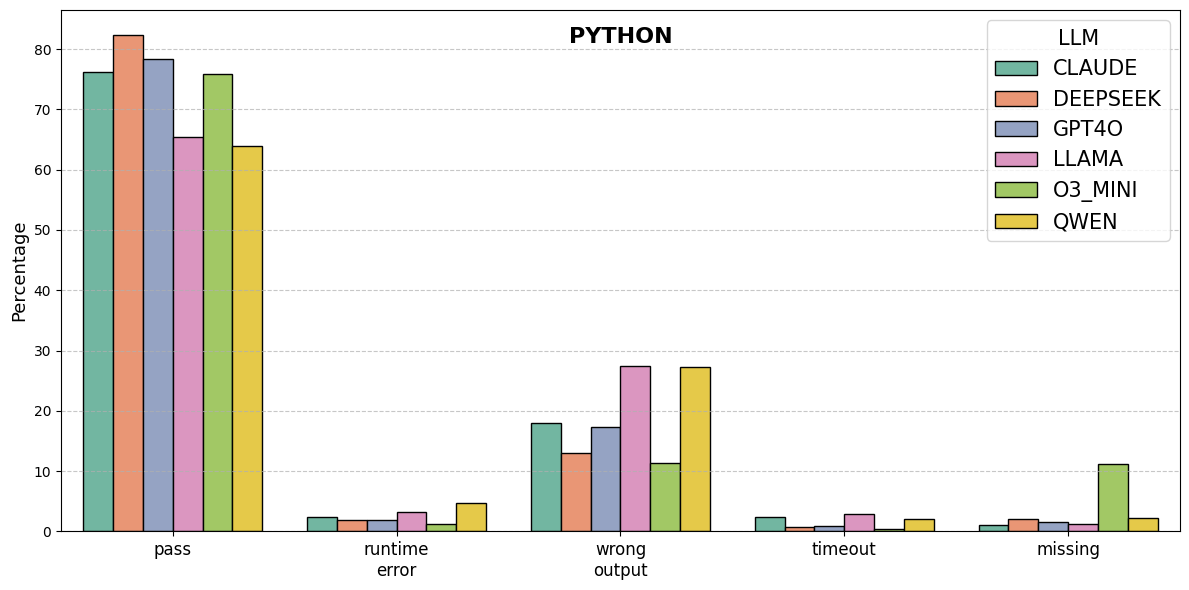}
\includegraphics[width=0.75\linewidth]{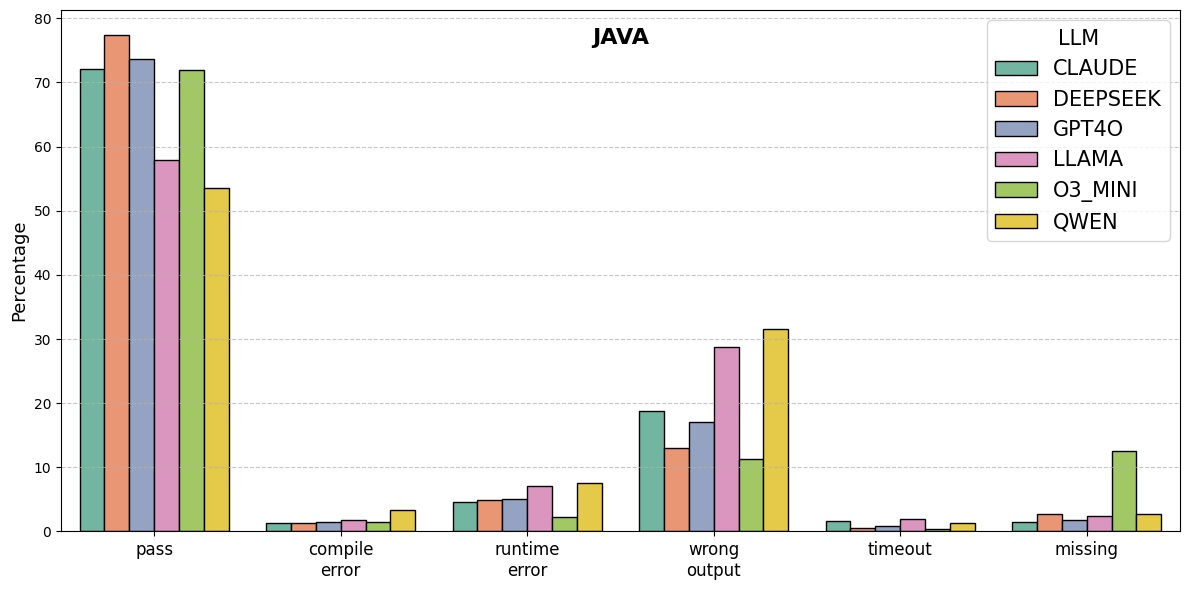}
\includegraphics[width=0.75\linewidth]{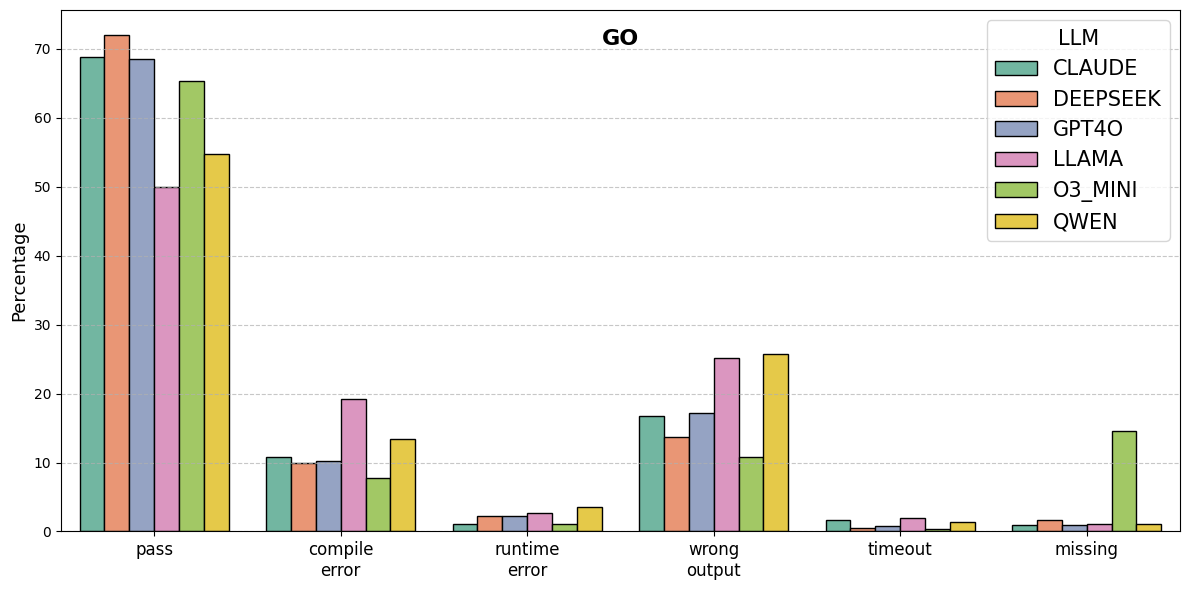}
\caption{Distribution of generated code outcomes across LLMs for each language. The y-axis shows the percentage of each outcome, with the sum of outcome categories for each LLM totaling 100\%.}
\label{fig:status_distribution}
\end{figure*}

\noindent
\textbf{Impact of LLMs.} First, we investigate the impact of LLMs in code accuracy  using all prompts and calculate the proportional distribution of the outcomes for each LLM separately.   \textbf{Figure \ref{fig:status_distribution}} illustrates these results across the Python, Java, and Go programming languages.

\begin{comment}

\begin{figure*}[!htb]
    \centering
    \begin{subfigure}[!htb]{0.9\linewidth}
    \includegraphics[width=0.9\linewidth]{py_fig/py_llm_durum2.png}
    \end{subfigure}
    \begin{subfigure}[!htb]{0.9\linewidth}
      \includegraphics[width=0.9\linewidth]{py_fig/java_llm_durum2.png}
    \end{subfigure}
    \begin{subfigure}{0.9\linewidth}
     \includegraphics[width=0.9\linewidth]{py_fig/go_llm_durum2.png}
    \end{subfigure}
    \caption{Distribution of generated code outcomes across LLMs for each language. The y-axis shows the percentage of each outcome, with the sum of outcome categories for each LLM totaling 100\%.}
    \label{fig:status_distribution}
\end{figure*}
\end{comment}

%The analysis of Python code outcomes across the evaluated LLMs reveals notable differences in performance and error tendencies as shown in \textbf{Figure \ref{fig:py_status_distribution}}. 
%among the models, DeepSeek achieves the highest success rate in all programming languages. In addition, 

Our observations are as follows. First, the ranking of models by success rate is consistent across both Python and Java. Specifically, DeepSeek achieves the highest success rate, followed by GPT-4o, while O3-mini and Claude exhibit nearly identical performance. QWen ranks lowest among the evaluated models. In the case of the Go programming language, the ranking differs slightly: DeepSeek again leads, but GPT-4o and Claude share the second position, both outperforming O3-mini. Interestingly, in Go, QWen outperforms LLaMA, a trend not observed in Java and Python. Notably, across all models, the pass rates for Java are lower than those for Python.

In failed cases (i.e., compile-time errors, runtime errors, incorrect outputs, and timeouts), we find that incorrect outputs are the most common failure case across all languages. This indicates that the models are generally capable of generating executable code, albeit often with algorithmic flaws. In addition, %we report compilation error rates only for Go and Java, as Python does not require compilation. 
we observe that LLMs tend to produce more compilation errors in Go than in Java. Furthermore, LLaMA and QWen exhibit significantly higher compilation error rates compared to the other models.

%followed by GPT-4o, O3 Mini, and Claude. %In contrast,  Llama and Qwen exhibited lower pass rates.
%A closer look at failure modes highlights varying patterns across the models. Qwen and Llama produce the highest proportions of runtime errors and incorrect outputs. %, indicating a tendency to generate code that fails during execution. These two models were also more prone to generating incorrect outputs—cases where the code executes but yields the wrong result—reflecting broader challenges in both execution reliability and solution correctness.

Interestingly, O3-mini exhibits the highest proportion of missing cases across all scenarios. Our manual inspection reveals that O3-mini frequently fails to return any response to a given prompt, often due to extended response times associated with its reasoning process. However, it is noteworthy that O3-mini demonstrates the lowest rates of runtime errors, incorrect outputs, and timeouts. When considering only the cases in which the model produces code (i.e., excluding missing responses), O3-mini achieves the highest success rate. These findings suggest that, although O3-mini may occasionally fail to respond, its reasoning capabilities allow it to generate highly accurate solutions—albeit at the expense of speed or response consistency.

\begin{figure*}[!htb]
\centering
\includegraphics[width=0.95
\linewidth]{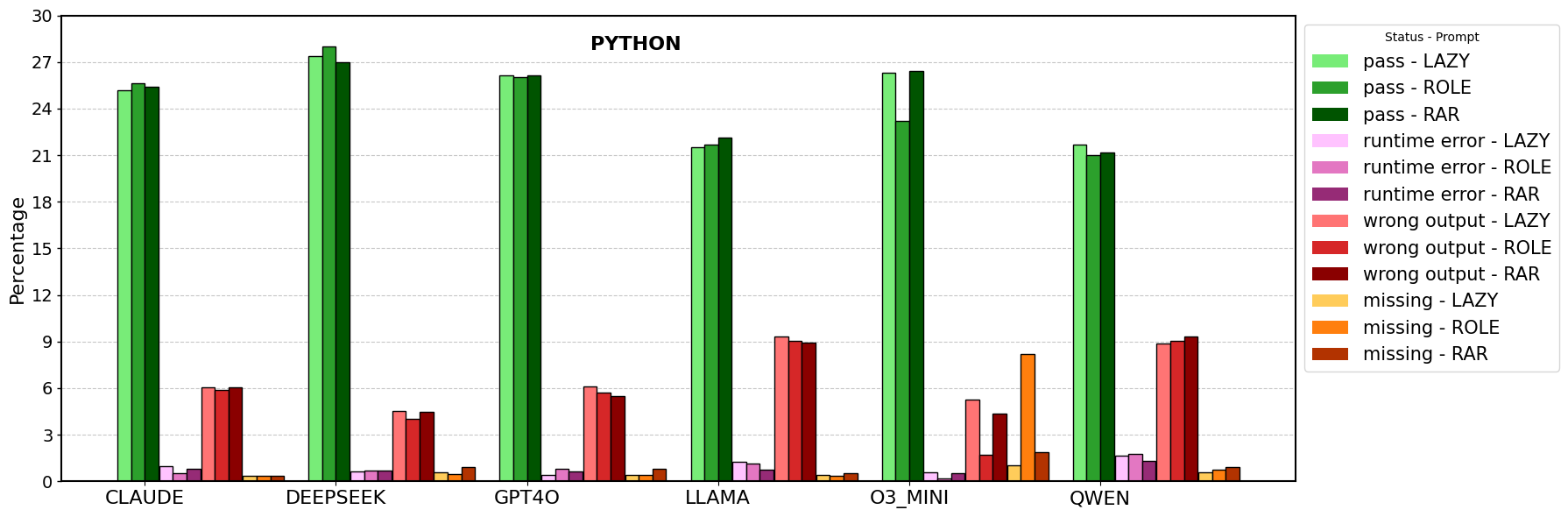}
\includegraphics[width=0.95\linewidth]{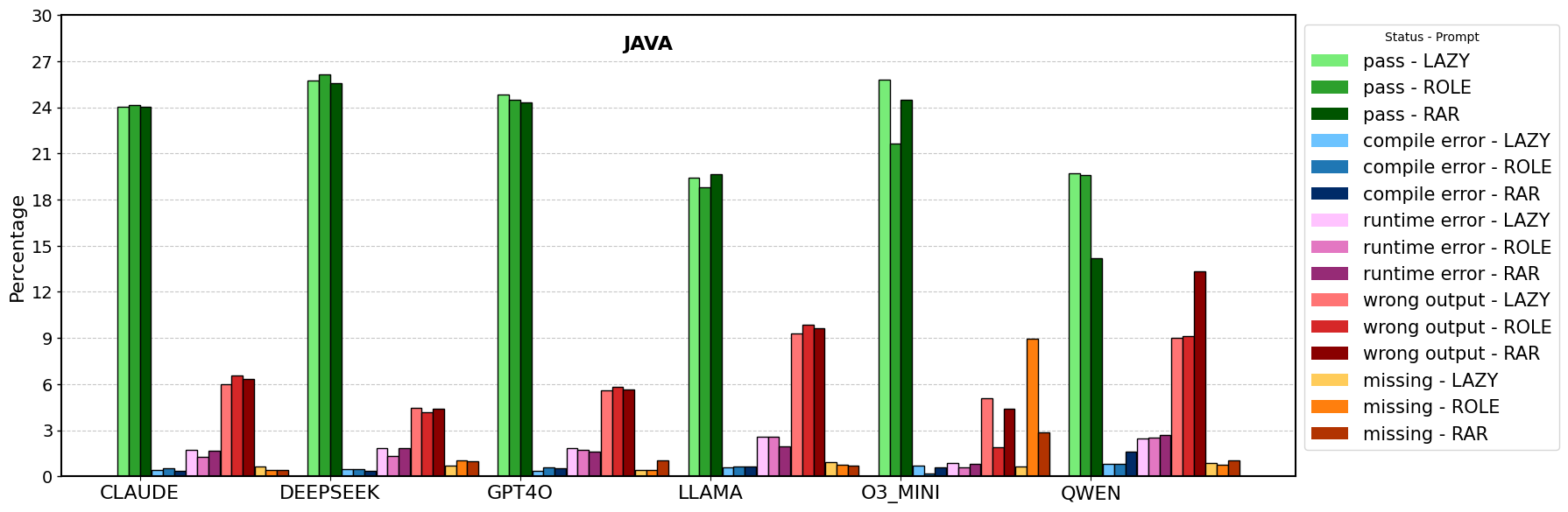}
\includegraphics[width=0.95\linewidth]{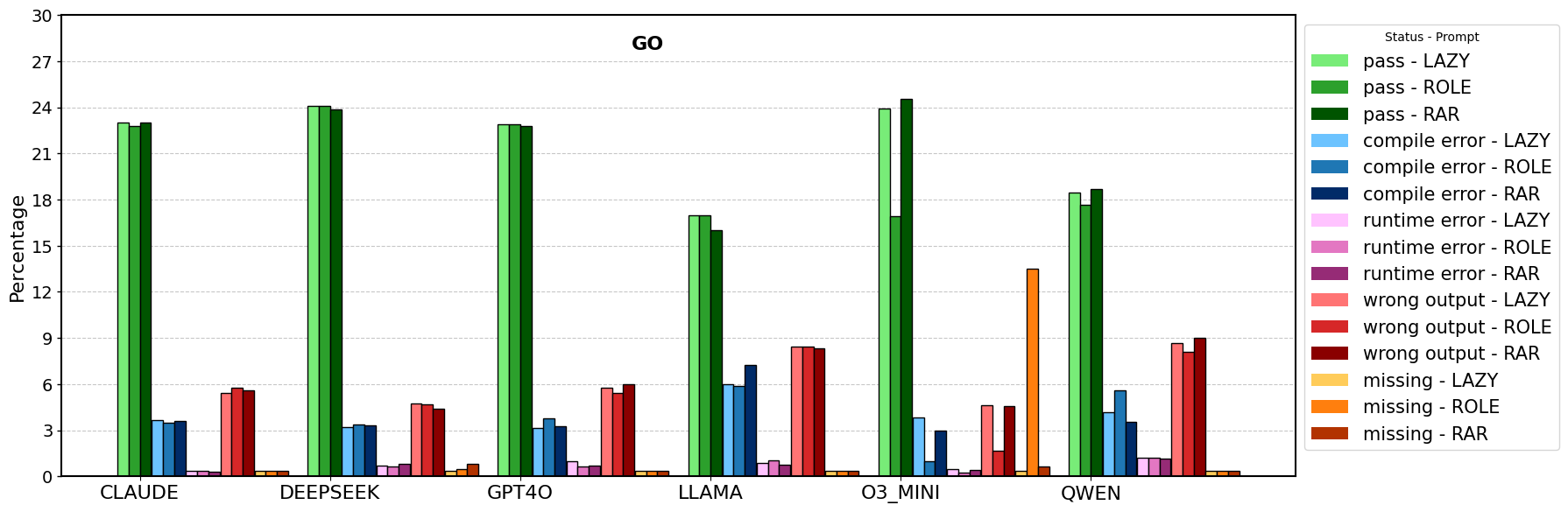}
\caption{LLM-wise distribution of output statuses per prompt type. Unlike the previous analysis, which reported overall outcome rates per model, this figure breaks down each outcome category into three bars representing the contributions of the different prompts used. Timeout errors are excluded for clarity, so totals may not sum to 100\%.}
\label{fig:prompt_language_status}
\end{figure*}

\noindent
\textbf{Impact of Prompting Strategies.}  We  now focus on the  impact of prompt variation on code generation outcomes. Therefore, we calculate the proportion of each outcome  separately for each prompt. 
%covering both successful and failed code executions across different error categories.
\textbf{Figure~\ref{fig:prompt_language_status}} presents the distribution of code generation outcomes across different prompt types for each programming language (Python, Java, and Go).

%To avoid visual clutter caused by the large number of outcome categories and prompt variations, the ‘timeout’ category was excluded from this visualization. As a result, the outcome percentages for each model in this figure may not always sum precisely to 100\%.

%When aggregating the pass rates across languages, we observe that all LLMs consistently achieve their highest success rates in Python, followed by Java, with the lowest performance in Go. This pattern, which was already evident in the language-specific analyses, is further confirmed by the three prompt-based outcome tables. These results suggest that, within this dataset, LLMs are generally more proficient at generating correct Python code compared to Java and Go, likely reflecting both the models’ training distributions and the relative complexity of the languages.

We observe that the impact of prompt variation differs across LLMs and programming languages. No single prompting strategy consistently leads to higher or lower performance across all models. Furthermore, with the exception of O3-mini, no prompting strategy consistently results in the highest or lowest success rate for a given LLM. For example, while the \textit{Rephrase and Respond} strategy yields the highest pass rates in Python and Java for LLaMA, it produces the lowest pass rates in Go.

In the case of O3-mini, we find that the Role prompting strategy leads to a significantly higher number of missing responses across all languages. This suggests that Role prompts may yield deeper or more complex reasoning, which in turn increases response latency and can prevent the model from generating code within the expected time frame.

%The tables also highlight the influence of prompt variation on code correctness. Prompt design clearly affects whether the models produce code that successfully executes or fails. This effect is particularly pronounced in models such as O3 Mini, Qwen, and Llama, where noticeable differences in pass rates and error distributions emerge across prompt types.

%For O3 Mini, across all programming languages, the ‘role’ prompting consistently yields the lowest pass rates. However, this same prompt type also results in the lowest proportions of compile errors, runtime errors, and incorrect outputs. The high number of ‘missing’ cases under role prompting—where the model frequently fails to generate any response—accounts for this discrepancy. Manual inspection confirmed that when O3 Mini does respond under role prompting, it tends to produce more accurate code, suggesting that while this prompting style increases the likelihood of non-response, it also enhances the correctness of the generated code when a response is given.

%For Qwen, prompt variation has a particularly visible impact on Java code generation. The ‘rar’ prompting led to the lowest pass rates and the highest proportions of all error types in Java, demonstrating the model’s sensitivity to prompt phrasing in this language.

Overall,  our results demonstrate the importance of using diverse prompt types and multiple programming languages  to reiably evaluate the code generation capabilities of LLMs.

\section{Experiments}\label{sec_exp}

 In this section, we explain our experimental setup and present benchmark results for state-of-the-art  AI-generated code detection models on  \corpus{}. %While the data preprocessing steps are detailed in Section 3, here we focus on our training, validation, and test set splitting strategy, dataset analysis, and the embedding extraction process. We then evaluate three distinct experimental scenarios: overall performance analysis, cross-LLM evaluation, and cross-programming language assessment.

\subsection{Experimental Setup}

\noindent
\textbf{Dataset.} We divided the 800 problems from CodeNet into training (80\%), validation (10\%), and test (10\%) sets to ensure that codes for the same problem never appears in both train and test sets. %This approach prevents potential data leakage that could occur if similar solutions for the same problem appeared in both training and evaluation sets.
Furthermore, to ensure a more balanced distribution across the splits, we considered the 'score' values provided by CodeNet for each problem. We then distributed the problems among the training, validation, and test sets while maintaining score-based balance. 
%factored in these problem scores when assigning problems to the training, validation, and test sets. This additional consideration helps maintain a representative mix of problems across varying levels of complexity, further strengthening the reliability of our evaluation.

We conduct experiments for our three code generation scenarios: Scenario\textsubscript{Scratch}, Scenario\textsubscript{Runtime}, and Scenario\textsubscript{Output}. For each specific scenario, we utilize all human-authored codes and the codes generated for that scenario, adhering to the designated data splits across training, testing, and validation sets. 
%Since the AI generated codes are more than the human-authored ones in our dataset (See Table \ref{Tab_stats}), we use all human-authored code samples in all experiments while we use different subsets of AI generated code samples based on the needs of the experiments. 

%In our experiments,  we use all preprocessed human samples across all scenarios 
%given the limited data availability,. For LLM-generated code, we maintain the problem-based split and organize the data according to our three  scenarios: i) generating code from scratch, ii) fixing runtime errors, and iii) correcting incorrect outputs.

\noindent
\textbf{AI Generated Code Detection Models.}
In our experiments, we use the following state-of-the-art models:

\begin{itemize}
    \item \textbf{SVM\textsubscript{Ada}.} We train an SVM model using OpenAI's text-embedding-ada-002\footnote{\url{https://openai.com/index/new-and-improved-embedding-model/}} code embedding model. We choose Ada embedding models because of its success in producing high-quality code representations and popularity in the literature \citep{oedingen2024chatgpt}. 
    
    \item \textbf{SVM\textsubscript{T5+}.} We train an SVM model using Salesforce's CodeT5+\footnote{\url{https://huggingface.co/Salesforce/codet5p-110m-embedding}} embedding model because of its frequent use in similar tasks such as code writer detection and feature extraction \citep{pham2024magecode, suh2024empirical, gurioli2024you}. 
    %We utilize Salesforce/codet5p-110m-embedding\footnote{\url{https://huggingface.co/Salesforce/codet5p-110m-embedding}} to generate 256-dimensional embeddings specifically designed for code semantics.
    \item \textbf{CodeBERTa.} We fine-tuned CodeBERTa\footnote{\url{https://huggingface.co/huggingface/CodeBERTa-small-v1}}, which is a distilled version of CodeBERT \citep{feng2020codebert} that has been widely used in several prior works  for feature extraction and detector design \citep{xu2024detecting, bulla2024ex,nguyen2024gptsniffer}.
\end{itemize}

%this study, we chose two state-of-the-art code embedding models: i) Ada\footnote{\url{https://openai.com/index/new-and-improved-embedding-model/}} (OpenAI’s text-embedding-ada-002) and ii) T5+\footnote{\url{https://huggingface.co/Salesforce/codet5p-110m-embedding}}, due to their ability to produce high-quality code representations and popularity in the literature (e.g., \cite{oedingen2024chatgpt,pham2024magecode, suh2024empirical, gurioli2024you}. %We did not choose OpenAI’s text-embedding-v3 because Ada-embedding-v002 is more widely used in the literature on our topic \cite{oedingen2024chatgpt}. 
%T5+ was chosen because of its frequent use in similar tasks such as code writer detection and feature extraction \cite{pham2024magecode, suh2024empirical, gurioli2024you}. 

\noindent
\textbf{Implementation.}  We used Scikit's SVC library\footnote{\url{https://scikit-learn.org/stable/modules/generated/sklearn.svm.SVC.html}} for SVM\textsubscript{Ada} and SVM\textsubscript{T5+} models. We tuned the hyper-parameters using a grid search on the validation set. In particular, we set gamma to \textit{scale} since it is the best parameter in all scenarios, and we set kernel to \textit{RBF} also gave the best results in most of the kernel parameters. The C value varies depending on the scenario. We use Trainer\footnote{\url{https://huggingface.co/docs/transformers/main_classes/trainer}} library to tune the hyperparemeters of CodeBERTa using the validation set. We set epoch to 3 and batch size to 8 accordingly. 
The input codes are truncated to fit the embedding models' token limits when necessary.

We spent approximately \$150 for APIs of models during code generations. We also used a combination of personal GPUs (RTX3080 and RTX4096) and cloud services (A100 GPU on Google Colab) for experiments.

\subsection{Experimental Results}
In this section, we provide benchmark results for state-of-the-art models for Python, Go, and Java languages in our three different scenarios using  \corpus{}. We also assess their performance in cross-model and  cross-language setups.

\begin{table*}[!htb]
\small
\setlength\tabcolsep{3pt}
\centering
  \begin{tabular}{ | l | l | l |l |l ||l |l |l ||l |l |l  |l |l |l | }
    \hline
    & &\multicolumn{3}{c||}{\textbf{Java}} & \multicolumn{3}{c||}{\textbf{Python}} & \multicolumn{3}{c|}{\textbf{Go}}  \\ \hline
  & & SVM\textsubscript{Ada} & SVM\textsubscript{T5+} & CodeBERTa & SVM\textsubscript{Ada} & SVM\textsubscript{T5+}  & CodeBERTa  & SVM\textsubscript{Ada} & SVM\textsubscript{T5+} & CodeBERTa    \\ \hline
  \multirow{4}{*}{\rotatebox{90}{Scratch}} &  A &  0.9759 & 0.9452 & \textbf{0.9798} & 0.9177 & 0.8781 & \textbf{0.9441} & 0.9363 & 0.9108 & \textbf{0.9446} \\ 
  &  P &  \textbf{0.9764} & 0.9403 & 0.9726 & \textbf{0.9546} & 0.9346 & 0.9529 & \textbf{0.9643} & 0.9281 & 0.9465 \\ 
  &  R &  0.9792 & 0.9599 & \textbf{0.9907} & 0.8963 & 0.8423 & \textbf{0.9474} & 0.9291 & 0.9241 & \textbf{0.9628} \\ 
   & $F_1$ &  0.9778 & 0.95 & \textbf{0.9815} & 0.9245 & 0.8861 & \textbf{0.9501} & \textbf{0.9664} & 0.9261 & 0.9546 \\ \hline \hline
  \multirow{4}{*}{\rotatebox{90}{Runtime}} & A & 0.7925 & 0.7382 & \textbf{0.8038} & \textbf{0.7256} & 0.6568 & 0.7252 & \textbf{0.8694}  & 0.6401 & 0.7937\\ 
  & P & 0.8486 & 0.7635 & \textbf{0.8614} & 0.8414 & 0.7320 & \textbf{0.8459} & \textbf{0.8845}  & 0.6525 & 0.8520 \\ 
  & R & 0.7498 & 0.7469 & \textbf{0.7592} & \textbf{0.6306} & 0.6142 & 0.6249 & \textbf{0.8429}  & 0.5642 & 0.6995 \\ 
  &  $F_1$ & 0.7962 & 0.7551 & \textbf{0.8071} & \textbf{0.7209} & 0.6680 & 0.7188 & \textbf{0.8632}  & 0.6052 & 0.7683 \\ \hline \hline
  \multirow{4}{*}{\rotatebox{90}{Output}} & A & 0.7515 & 0.7180 & \textbf{0.7981} & 0.7273 & 0.6937 & \textbf{0.7694} & \textbf{0.8126} & 0.6931 & 0.7950 \\ 
  &  P & 0.8357 & 0.7463 & \textbf{0.8465} & \textbf{0.8301} & 0.7486 & 0.8079 & \textbf{0.8844} & 0.6912 & \textbf{0.8844} \\ 
  &  R    & 0.6746 & 0.7276 & \textbf{0.7670} & 0.6462 & 0.6833 & \textbf{0.7727} & 0.7898 & \textbf{0.8798} & 0.7559 \\ 
  &   $F_1$        & 0.7465 & 0.7368 & \textbf{0.8048} & 0.7267 & 0.7145 & \textbf{0.7899} & \textbf{0.8344} & 0.7742 & 0.8151 \\ \hline
  \end{tabular}
  \caption{The classification performance of modes for each scenario and programming language. The highest performing result for each case is written in \textbf{bold}. A: Accuracy, P: Precision, R: Recall, Scratch: Scenario\textsubscript{Scratch}, Runtime: Scenario\textsubscript{Runtime}, Output: Scenario\textsubscript{Output}.}
  \label{tab_result_overall}
\end{table*}

\subsubsection{Multi-LLM and Multi-language Scenario-specific Training} 
In this experiment, we separately train the models  for each scenario using all human-authored codes and the codes generated for that specific scenario in the training set, covering  multiple LLMs and  the three programming languages.
%We trained separate models for each usage scenario (code generation, runtime error fixing, and output correction) using our three embedding approaches. For each scenario, we evaluated the models across all language-LLM combinations in our dataset, covering Python, Java, and Go languages, and seven different sources (Qwen, Llama, DEEPSEEK, Claude, GPT4O, O3\_MINI, and human-authored code).
We calculate the classification performance of the trained models on the test set for each programming language separately. The results are shown in \textbf{Table \ref{tab_result_overall}}.

We observe that CodeBERTa achieves the highest $F_1$ scores across all scenarios for Java and in most cases for Python, whereas SVM\textsubscript{Ada} consistently has the highest $F_1$ scores for Go. Notably, while all models have high performance in Scenario\textsubscript{Scratch}, their effectiveness declines in the other scenarios, which is expected given that these scenarios require the models to fix a given code, instead of generating from scratch. Furthermore, the models have the lowest average performance for Python and the highest for Java. Overall, our findings show the importance of selecting detection models based on both the programming language and the LLM usage scenario.

To analyze performance of detection models for each LLM, we report their accuracy separately in \textbf{Table \ref{tab_create}}, \textbf{Table \ref{tab_runtime}}, and \textbf{Table \ref{tab_correction}} for Scenario\textsubscript{Scratch}, Scenario\textsubscript{Runtime}, and Scenario\textsubscript{Output}, respectively.
Our observations are as follows. Firstly, the best-performing model for detecting human-authored code often differs from the best-performing model for detecting LLM-generated code. For example, in Scenario\textsubscript{Scratch}, CodeBERTa achieves the highest accuracy in identifying LLM-generated code, whereas SVM\textsubscript{Ada} performs best on human-authored code.  Similarly, while  SVM\textsubscript{T5+} has the highest accuracy in detecting Go code in Scenario\textsubscript{Output} for almost all LLMs, its performance significantly decreases in identifying human-authored code. 

Secondly,  although CodeBERTa demonstrates higher average performance, model accuracy in detecting LLM-generated code varies across scenarios and programming languages. For example, in the Go language, the model with the highest accuracy for LLM-generated code differs across scenarios.

Thirdly, the accuracy of models in detecting code from a specific LLM can vary significantly depending on the scenario. For instance, the models achieve the highest accuracy for Qwen-generated code in Scenario\textsubscript{Scratch}, whereas in Scenario\textsubscript{Runtime}, Qwen-generated code becomes the most difficult to detect.

Lastly, in Scenario\textsubscript{Scratch}, where models are better able to reflect their own coding style compared to other scenarios, the highest performance is observed for code generated by Qwen and GPT-4o. In contrast, performance is generally lower for code generated by Llama and Claude, suggesting that these LLMs generate code that more closely resembles human-authored code.
Overall, our results show the importance of carefully selecting detection models based on both the scenario and language.

%Similar to the results for Scenario\textsubscript{Scratch} in Table \ref{tab_create}, SVM\textsubscript{Ada}  outperforms others in detecting LLM-generated codes but CodeBERTa is superior to others in detecting human-authored code. SVM\textsubscript{Ada}  is particularly successful in detecting codes generated by DeepSeek V3, Claude, and GPT-4o. 
%In both scenarios,  models have the lowest performance in detecting Qwen-generated codes, with the exception of the Go programming language in Scenario\textsubscript{Output}.  

%Notably, performance is particularly strong for Go, with several LLMs achieving perfect detection rates. Tables \ref{tab:runtime} and \ref{tab:correction} show similar patterns for runtime error fixing and output correction scenarios, though with slightly lower overall performance.

\begin{table*}[!htb]
\small
\setlength\tabcolsep{2pt}
\centering
  \begin{tabular}{ | l | l |l |l ||l |l |l ||l |l |l  |l |l |l | }
    \hline
    &\multicolumn{3}{c||}{\textbf{Java}} & \multicolumn{3}{c||}{\textbf{Python}} & \multicolumn{3}{c|}{\textbf{Go}}  \\ \hline
   & SVM\textsubscript{Ada} & SVM\textsubscript{T5+} & CodeBERTa & SVM\textsubscript{Ada} & SVM\textsubscript{T5+}  & CodeBERTa  & SVM\textsubscript{Ada} & SVM\textsubscript{T5+} & CodeBERTa    \\ \hline
  Qwen  & \textbf{1.0} & 0.9873 & \textbf{1.0}  & 0.9667& 0.8625 & \textbf{0.9833}  & 0.9750 & 0.9625& \textbf{0.9917}  \\ \hline
   Llama   & 0.9708 & 0.9875 &\textbf{0.9958}  & 0.8417 & 0.8125 & \textbf{0.9375}  & 0.8708 & 0.8917& \textbf{0.9125} \\ \hline
    
    DeepSeek V3  & \textbf{0.9873} & 0.9662 & 0.\textbf{9873}  & 0.9042 & 0.8375 & \textbf{0.9375}  & 0.9289& 0.9163 & \textbf{0.9623}  \\ \hline
    
    Claude  & 0.9792& 0.95 & \textbf{0.9917} & 0.85 & 0.8042 & \textbf{0.90} & 0.9083 & 0.90 & \textbf{0.9625} \\ \hline
    
    GPT-4o   & 0.9828 & 0.9828 & \textbf{0.9957} & 0.9244 & 0.8992 &\textbf{0.9958}  & 0.9542 & 0.9583 & \textbf{0.9792}  \\ \hline
    
    o3 - mini   & 0.9522 & 0.8756 & \textbf{0.9713} & 0.8905 & 0.8381 & \textbf{0.9286} & 0.9394 & 0.9141& \textbf{0.9697} \\ \hline
    
    Human   & \textbf{0.9719} & 0.9277 & 0.9668 & \textbf{0.9452}& 0.9242 & 0.9397 & \textbf{0.9474} & 0.8904&0.9167  \\ \hline 

  \end{tabular}
  \caption{The accuracy for each case in the "Generating the code from scratch" scenario. The highest performing result for each case is written in \textbf{bold}. }
  \label{tab_create}
\end{table*}

\begin{table*}[!htb]
\small
\setlength\tabcolsep{2pt}
\centering
  \begin{tabular}{ | l | l |l |l || l |l |l || l |l |l | }
    \hline
    &\multicolumn{3}{c||}{\textbf{Java}} & \multicolumn{3}{c||}{\textbf{Python}} & \multicolumn{3}{c|}{\textbf{Go}}  \\ \hline
    & SVM\textsubscript{Ada} & SVM\textsubscript{T5+} & CodeBERTa & SVM\textsubscript{Ada} & SVM\textsubscript{T5+}  & CodeBERTa  & SVM\textsubscript{Ada} & SVM\textsubscript{T5+} & CodeBERTa    \\ \hline  
  Qwen  & \textbf{0.7026} & 0.6121 & 0.6379  & 0.4723 & 0.4638 & \textbf{0.4809}  & \textbf{0.8163} & 0.4762 & 0.5782  \\ \hline
   Llama    & 0.6114 &\textbf{0.7118} & 0.7031  & 0.5272 & \textbf{0.6025} & 0.5565  & \textbf{0.7143} & 0.5510 & 0.6122 \\ \hline
    DeepSeek V3  &0.7373 & 0.75 & \textbf{0.7797}  & 0.6483 & 0.6525 & \textbf{0.6864}  & \textbf{0.8231} & 0.4966 & 0.6667  \\ \hline
    Claude  & 0.8529 & 0.8613 & \textbf{0.8824} & 0.6917 & 0.6750 & \textbf{0.6958} & \textbf{0.9320} & 0.6667 & 0.8027 \\ \hline
    GPT-4o   & \textbf{0.8458} & 0.7875 & \textbf{0.8458} & \textbf{0.7490} & 0.7280 & 0.7071  & \textbf{0.8912} & 0.6735 & 0.8299  \\ \hline
    o3 - mini   & 0.7404 & \textbf{0.7548} & 0.6923 & \textbf{0.6991} & 0.5556 & 0.6204  & \textbf{0.8832} & 0.5182 & 0.7080 \\ \hline
    Human   & 0.8427 & 0.7279 & \textbf{0.8563} & 0.8475 & 0.7114 & \textbf{0.8539} & \textbf{0.8947} & 0.7127 & 0.8838  \\ \hline \hline
  \end{tabular}
  \caption{The accuracy and other metrics for each case in fixing the runtime error scenario. The highest performing result for each case is written in \textbf{bold}.  } 
  \label{tab_runtime}
\end{table*}

\begin{table*}[!htb]
\small
\setlength\tabcolsep{2pt}
\centering
  \begin{tabular}{ | l | l | l | l || l | l | l || l | l | l | }
    \hline
    &\multicolumn{3}{c||}{\textbf{Java}} & \multicolumn{3}{c||}{\textbf{Python}} & \multicolumn{3}{c|}{\textbf{Go}}  \\ \hline
   & SVM\textsubscript{Ada} & SVM\textsubscript{T5+} & CodeBERTa  & SVM\textsubscript{Ada} & SVM\textsubscript{T5+}  & CodeBERTa   & SVM\textsubscript{Ada} & SVM\textsubscript{T5+} & CodeBERTa    \\ \hline

  Qwen  & 0.5714 & 0.5840 & \textbf{0.6513}  & 0.5 & 0.5085 & \textbf{0.6483}  & \textbf{0.8803} & \textbf{0.8803} & 0.8162  \\ \hline
  
   Llama   & 0.5401 & 0.6287 & \textbf{0.6751}  & 0.4667 & \textbf{0.5625} & 0.5333  & 0.6453 & \textbf{0.8205} & 0.6026 \\ \hline
   
    DeepSeek V3  & 0.6444 & 0.7364 & \textbf{0.7886}  & 0.7185 & 0.7773 & \textbf{0.8613}  & 0.7265    & \textbf{0.9017} & 0.6752  \\ \hline
    
    Claude  & 0.8452 & 0.8787 & \textbf{0.8954}  & 0.7197 & 0.7197 & \textbf{0.8452}  & 0.8974 & \textbf{0.9402} & 0.8932 \\ \hline
    
    GPT-4o   & 0.75 & 0.7875 & \textbf{0.8417}  & 0.7741 & 
    0.7908 & \textbf{0.8954}  & 0.765    & \textbf{0.9103} & 0.7521 \\ \hline
    
    o3 - mini   & 0.698 & \textbf{0.7525} & 0.7475  & 0.7053 & 0.7488 & \textbf{0.8647}  & \textbf{0.8833} & 0.8118 & 0.8065 \\ \hline
    
    Human   & \textbf{0.8427} & 0.7066 & 0.835  & \textbf{0.8311} & 0.7068 & 0.7653  & 0.8465 & 0.4156 & \textbf{0.8531} \\ \hline 
  \end{tabular}
  \caption{The accuracy for each case in correcting the output scenario. The highest performing result for each case is written in \textbf{bold}. }
  \label{tab_correction}
\end{table*}

% bir tane devasa modelimiz olsun, tüm diller ile eğitilmiş bir model. 
% tüm train ile eğitilmiş modelimiz olacak.
% bir de her programlama dili için ayrı bir model - vaktimiz olursa

% tüm test kümesinde sonuç alacağız. 
% her case'deki başarısı nedir? 3 programlama dili ve 3 generation tipi (from scratch, fix runtime error, wrong fix)

% -

\subsubsection{Cross LLM Performance.} 
In this experiment, we conduct \textit{leave-one-LLM-out} experiment to evaluate  models' performance in detecting  code samples generated by LLMs not seen in the training set. 
%ability to detect code from unseen LLMs, we conducted leave-one-out experiments focusing on the code generation scenario. 
We focus exclusively on Scenario\textsubscript{Scratch}, as it more accurately reflects the coding style and capabilities of LLMs. We  train each model using training data from all LLMs except one and then evaluate its performance  on the held-out LLM's test data combined with all human-authored code samples in the test data. %This approach helps assess how well models generalize to new LLMs.

\textbf{Table \ref{tab:cross_llm}} presents  $F_1$ scores for each LLM we target.  CodeBERTa  achieves the highest $F_1$ score in four (out of six) cases. 
In terms of performance across LLMs, we observe substantial variation in model accuracy, suggesting that different LLMs exhibit distinct coding styles. Notably, the models perform worst on code generated by Llama and o3-mini, while achieving their highest performance on GPT-4o-generated code (e.g., CodeBERTa reaches an $F_1$ score of 0.9240).

%models have the lowest performance in detecting o3-mini generated code samples. This might be because o3-mini is the only reasoning-based LLM we use.

%F1 scores above 0.93 for most LLMs, with particularly strong performance when detecting GPT4o (0.9783) and Qwen (0.9750) generated code. While CodeT5 and CodeBERT embeddings show slightly lower performance, they still maintain F1 scores above 0.85 in most cases, demonstrating the general effectiveness of our approach for unseen LLMs.

% bir tane devasa modelimiz olsun, tüm diller ile eğitilmiş bir model. 
% 7 farklı sonuç alacağız. 
% leave-one-out
% sadece generate from scracth - train ve test verisinde sadece bunlar olsun
% 3 prompt'u da kullanalım
% her llm için yaklaşık 800x3 veri var. 
% human'ın test kümesinde  benzer sayıda sample alalım (bu sabit olacak her 6 deneyde de)
% trainden bir LLM'e ait sonuçları çıkarıp sonra modeli train edip, sonra test kümesinde o llm'e ait code'lar ve X\_human

\begin{table}[!htb]
\small
\centering
\setlength\tabcolsep{2pt}
  \begin{tabular}{ | l | l || l |c |c | c | }
    \hline
    Train & Test  & SVM\textsubscript{Ada} & SVM\textsubscript{T5+} & CodeBERTa \\ \hline
  w/o Qwen & Qwen + H.  & 0.8968 & 0.7895 & \textbf{0.9085}  \\ \hline
  w/o  Llama & Llama + H.   & 0.7182 & \textbf{0.7259} &0.6650  \\ \hline
   w/o  DeepSeek V3 & DeepSeek V3  + H. & 0.855& 0.7929&\textbf{0.8693}  \\ \hline
  w/o   Claude  & Claude  + H. & 0.8045& 0.7373& \textbf{0.8557} \\ \hline
  w/o   GPT-4o & GPT-4o  + H.   & 0.8803& 0.8106& \textbf{0.9240}  \\ \hline
  w/o   o3-mini  & o3-mini    + H.  & \textbf{0.7679} & 0.6997& 0.6969  \\ \hline
  \end{tabular}
  \caption{ F1 score of baseline systems in cross model setup for generating code from scratch scenario. H: Human-authoted data.  %In each case, we use all training data except the data of the corresponding LLM. In testing, we use the subset of test data that contains only human authored code and the code generated by the corresponding LLM.
  } 
  \label{tab:cross_llm}
\end{table}

\subsubsection{Cross Programming Language Performance. }
In our last experiment, we assess the models' performance in detecting AI generated codes for programming languages that do not exist in their train sets.
%to generalize across programming languages. 
In particular, for each language we target, we train the models with the code samples we have for the other languages and then evaluate their model on the test set of the corresponding language. Similar to the cross-LLM setup, we focus on Scenario\textsubscript{Scratch} and use all available LLMs and human-authored codes for the corresponding language  in the test set.

\textbf{Table \ref{tab:cross_lang}} presents the results for each model across programming languages. Compared to the results in Table \ref{tab_result_overall}, we observe that model performance in detecting AI-generated code  declines when the target programming language is not included in the training set.
Furthermore, detection performance drops significantly for the Go language. This may be attributed to the higher popularity and consequently greater representation of Python and Java in the training data of LLMs. Among the models, SVM\textsubscript{Ada} achieves the highest performance, while SVM\textsubscript{T5+}  yields the lowest score on average.
Overall, these findings demonstrate the importance of developing comprehensive datasets that span a diverse set of programming languages to improve the detection of AI-generated code.
%score when we testing with Python and Java languages . This might be because of LLMs better performance in these languages due to their higher  
%The results reveal interesting patterns in cross-language generalization. Models trained on Java and Go show strong performance when testing on Python (F1 = 0.8306 with CodeBERT), likely due to shared syntactic patterns. However, performance drops notably when testing on Go after training on Java and Python (F1 = 0.5060 with Ada embeddings), suggesting that Go's distinct syntax patterns make cross-language generalization more challenging.
%These results highlight both the strengths and limitations of our approach. While our models show strong performance within individual languages and reasonable generalization across similar languages, the detection of AI-generated code in syntactically distinct languages may require language-specific training data for optimal performance.
%Overall, we observe a significant drop in models' performance in cross-language setup. 

% 3 farklı sonuç alacağız. 
% leave-one-out
% sadece generate from scracth - train ve test verisinde sadece bunlar olsun
% 3 prompt'u da kullanalım
% tüm llm'ler kullanılacak
% human'ın test kümesinde  benzer sayıda sample alalım (bu sabit olacak her 3 deneyde de)
% trainden bir dile ait sonuçları çıkarıp sonra modeli train edip, sonra test kümesinde o dile ait code'lar üzerinden sonuç alınacak.

% F1, Prec, Recall, Accuracy

% (precision yuksek kalanlar kotu, f1 cok cok dusuk)
\begin{table}[!htb]
\small
\setlength\tabcolsep{3pt}
\centering
  \begin{tabular}{ | l | l || l |l |l | l | }
    \hline
    Train & Test  & SVM\textsubscript{Ada} & SVM\textsubscript{T5+} & CodeBERTa \\ \hline
 Java + Python  & Go  & \textbf{0.3085}&  0.1206&0.0712 \\ \hline
  Go + Python & Java    & 0.5856 & 0.4255& \textbf{0.7879} \\ \hline
  Java + Go & Python    & \textbf{0.8247} & 0.7885&  0.7655\\ \hline
  \end{tabular}
  \caption{ F1 score of models in cross programming language setup for generating code from scratch scenario. %In each case, we use all training data except the data of the corresponding LLM. In testing, we use the subset of test data that contains only human authored code and the code generated by the corresponding LLM.
  } 
  \label{tab:cross_lang}
\end{table}

%\subsubsection{Cross Prompt Performance }
%\hl{bu section doldurulmali}

\section{Limitations}\label{sec_limitations}
In our work, we developed a comprehensive dataset for detecting AI-generated code, covering six LLMs, three programming languages, three different prompting approaches, and three usage scenarios. We also provide benchmark results for three state-of-the-art detection models in various experimental setups. Despite its comprehensive coverage, it has the following limitations.

Firstly, the number of available models keeps increasing and the models become more advanced due to the fast developments in LLMs. Consequently, it is extremely challenging to cover all existing LLMs comprehensively. Therefore, our dataset requires periodic updates to incorporate emerging models and to phase out those that have become obsolete.

Secondly, detecting AI generated code is a highly active research field and there exist several models in the literature. However,  we could report results  for three models. Therefore, assessing other models will be an important extension of our work. %Furthermore, regarding the embedding models we use, we used only SVM models but other machine learning algorithm could be also examined.

%Thirdly, since our main objective is to detect the author of a code snippet, we did not evaluate whether the generated codes are  able to return the correct output for the given problem definitions.  These aspects can be considered as future work and may provide valuable insights for author detection analysis.

Lastly, while  being one of the most comprehensive dataset in the literature, the dataset can be always extended with more programming tasks, prompts, and programming languages. Our work do not cover all possible prompts that can be used for code generation.

\section{Ethical Considerations}\label{sec_ethic}
In our dataset, we do not share any personal information about the human authors of the codes. We re-organize and re-share subset of an existing dataset for human-authored code snippets to achieve systematic comparisons across models. Thus, we  do not think that any ethical issue might arise. Having said that, every technological tool might be used for good and bad purposes. We hope that our dataset will be used to develop effective AI-generated code detection models for good purposes. 

We would also like to clarify that AI tools were used only for grammar correction and writing improvement. No part of this paper was generated from scratch using AI.

\section{Conclusion}\label{sec_conc}
In this work, we introduce \corpus{} which covers multiple programming languages, LLMs, prompts, and usage scenarios. 
It contains 32,148 human-authored code samples and 121,271 AI-generated code samples. %We release \corpus{} to enable researchers to train their models and conduct systematic comparisons among different approaches.
%consisting of 32,190 human-authored and 122,888 AI-generated code samples. We share our dataset to help researchers to train their models and conduct systematic comparison with others.  
%a dataset for the AI-generated code detection problem, focusing on    Codestral, CodeLlama, and Gemini Flash 1.5 which are not well-studied in the literature. From the CodeNet dataset, we selected 317 problems and included five human submissions for each of the code status: Accepted, Runtime Error, and Wrong Answer, yielding 32,190 human-written code samples in total.  Next, we utilized the mentioned LLMs in three approaches: i) we used them to generate code based solely on the problem description  ii) we employed them to fix human-written code with runtime errors using the problem description, iii) we prompted them to correct human-written code that resulted in wrong answers, again with the problem description file. After filtering out improper outputs, we obtained a total of 155,078 AI generated code samples in total. 
In addition, we  provide analysis of LLM-generated code in terms of coding style and accuracy. Furthermore, we assess the performance of state-of-the-art  AI-generated code detection models using \corpus{}. In our experiments, we observe that models are highly capable of detecting codes generated from problem definition. However, their performance decreases in detecting AI-fixed code. We also observe that their performance varies across programming languages and reduces significantly in cross-language setup.

In the future, we plan to extend our dataset covering more programming languages, LLMs, and programming task. We also plan to cover more usage scenarios such as blended codes where LLMs are used to generate a portion of the code. Furthermore,  we will conduct a user study across students and software developers on how they use LLMs to generate code to identify realistic LLM usage scenarios. 

%the post-processing step, we were unable to verify whether the codes produced by the LLMs were complete, functional, or accurate. Specifically, we could not confirm if they generated the correct code or effectively fixed runtime errors or incorrect results in the sample codes, due to time constraints. We focused on whether the generated codes were written in Python, without considering their meaningfulness or accuracy. This can also be considered as potential future work on our dataset, focusing on evaluating the quality of LLMs or their ability to perform tasks and adapt based on prompts.

%\bibliographystyle{plainnat}
\bibliography{references}

\setcounter{table}{0}
\setcounter{figure}{0}
\renewcommand{\thetable}{A\arabic{table}}
\renewcommand{\thefigure}{A\arabic{figure}}

\section*{Appendix}

\begin{table*}[!htb]

\centering
  \begin{tabular}{ | p{3cm} |  l |  p{9cm} |}
    \hline
    All Scenarios & \textbf{System Prompt} &  You are a helpful prompt engineer. You will generate a new prompt from the given prompt for code generation. You make prompts more specific and detailed.\\ \hline
  Generate Code From Scratch & \textbf{User Prompt} & Paraphrase and expand this prompt: Write me a $<$programming language$>$ code to solve the following problem. \newline  Problem description: $<$problem description$>$  \newline 
    Just return the code without any explanation. Do not add ```  $<$Programming Language$>$ or  ```.
   \\ \midrule 
   Fix Runtime Error  & \textbf{User Prompt} & Paraphrase and expand this prompt: Fix the runtime error in the following $<$programming language$>$  code for the given problem. \newline Problem description: $<$problem description$>$  + \newline $<$programming language$>$  code to be fixed: \newline $<$Example solution$>$ \newline Just return the code without any explanation. Do not add ```  $<$programming language$>$ or ```. \\ \midrule 
   Correct the Output & \textbf{User Prompt} & Paraphrase and expand this prompt: Fix the following $<$programming language$>$  code that produces incorrect results for the given problem. \newline Problem description: $<$problem description$>$ \newline $<$programming language$>$  code to be fixed: \newline \{example solution\} \newline Just return the code without any explanation. Do not add \``` $<$programming language$>$  or \```. \\ \bottomrule 
  \end{tabular}
  \caption{  Prompts used to generate prompts for "Rephrase and Response" prompting approach. We use the same system prompt for all scenarios but user prompt changes based on the needs of coding scenario. } 
  \label{tab_prompts_for_prompts}
\end{table*}

\begin{table*}[!htb]
\small
\centering
  \begin{tabular}{ |  p{2cm} |  p{2cm} | p{10cm} |    }
    \toprule
    \multirow{3}{2cm}{Generate Code from Scratch} &\textbf{Lazy Prompting} & Write me a Python code to solve the following problem.  \newline Problem description: \{Problem description\} \newline  Just return the code without any explanation.. Do not add ``` Python  or ```. \\ 
    & \textbf{Role Specifying Prompting}  &  You are a helpful code assistant. Your language of choice is Python. You will generate the Python code for the given problem description. Remember, do not give any explanations, do not add ``` python or ```, just return the Python code block itself. Here is the problem description: \{Problem description\}  \\ 
   & \textbf{Rephrase and Response }  &  Develop a Python script that addresses the problem outlined below. The task is to design a complete and functional solution based solely on the provided problem statement. The problem description is as follows: \{Problem description\}. Your output should include only the Python code without any additional commentary or explanations. Please avoid enclosing your code within markdown-style code fencing markers (such as ```python``` or similar). \\ \hline
 
     \multirow{3}{2cm}{Fix  Runtime Error} & \textbf{Lazy Prompting} & Fix the runtime error in the following Python code for the given problem. \newline Problem description: \{Problem description\} \newline  Python  code to be fixed: \newline  \{Example solution\} \newline Just return the code without any explanation. Do not add ``` python or ```  \\ 
    & \textbf{Role Specifiying Prompting}  & You are an expert Python programmer that helps to fix the code for runtime errors. I will give you first the problem description, then the code that has the error. You will fix the Python code. Remember, do not give any explanations, do not add ``` python or ```, just return the Python code block itself. Here is the problem description: \{Problem description\} Here is the code that has the error: \{Example solution\}  \\ 
   & \textbf{Rephrase and Response }  & Your task is to analyze and correct a Python code snippet that is currently producing a runtime error. Review the accompanying problem description carefully to ensure that your modifications fully address the problem requirements. Once you have identified the source of the error, revise the code so that it runs correctly and meets the problem's specifications. Do not include any extra text or explanations—simply return the corrected Python code exactly as it should appear. Avoid using any formatting markers such as triple backticks. \newline \newline Problem description: \{Problem description\}  \newline Python code to be fixed: \{Example solution\} \\ \hline
   \multirow{3}{2cm}{Correct the Output} & \textbf{Lazy Prompting} & Fix the following Python code that produces incorrect results for the given problem. \newline Problem description: \{Problem description\}  \newline Python code to be fixed: \newline \{Example solution\} \newline Just return the code without any explanation. Do not add ```  python or ```.  \\ 
   & \textbf{Role Specifying Prompting}  &  You are an expert Python programmer that helps to fix the code results in incorrect answers. I will give you first the problem description, then the code that results in incorrect answers. You will fix the Python code. Remember, do not give any explanations, do not add ``` python or ```, just return the Python code block itself. Here is the problem description: \{Problem description\} Here is the code that results in wrong answers: \{Example solution\}  \\ 
   & \textbf{Rephrase and Response }  &  Review the provided Python code and modify it so that it correctly solves the problem described below. \newline \newline  Problem Description: \{Problem description\}\newline \newline  Python Code to Fix: \{Example solution\}\newline  \newline 
 Ensure your revised code resolves the errors and produces correct results based on the problem requirements. Return only the updated Python code without any extra explanation or markdown formatting like code block markers\\ \hline
  \end{tabular}
  \caption{Prompts we use to generate code in three different scenarios. We provide examples for Python code. For the other languages, we just replace it with the corresponding language name. For Rephrase and Response approach, we provide the results that o3-mini provides as each LLM generates a different prompt.} 
  \label{Tab_code_prompt}
\end{table*}

% \begin{table*}[!htb]
% \small
% \centering
%   \begin{tabular}{  |  p{5cm} | p{5cm} | p{5cm} |   }
%     \toprule
%     \textbf{Human-Written} & \textbf{Claude} & \textbf{DeepSeek} \\ \hline
%     Code & Code & Code \\ \midrule
%     \textbf{Llama}  & \textbf{GPT4O} & \textbf{O3 MINI} \\ \hline
%      Code & Code & Code \\ \midrule
% \textbf{Qwen}  & \textbf{DeepSeek-R1-Distill 70B}  & \textbf{DeepSeek-R1}  \\  \hline 
%    Code & Code & Code  \\ \bottomrule
%   \end{tabular}
%   \caption{ Human written and AI generated Java codes for the counting trailing zeros problem.} 
%   \label{Tab_code}
% \end{table*}

\begin{table*}[!htb]
  \tiny
  \centering
    \begin{tabular}{|p{4.5cm}|p{4.5cm}|p{4.5cm}|}
      \toprule
      \textbf{Human-Written} & \textbf{Claude} & \textbf{DeepSeek V3} \\ \hline
\begin{lstlisting}[style=pythonstyle]
n, a = map(int,input().split())
X = list(map(int,input().split()))

X.sort()
ii = 0
while ii < n and X[ii] < a:
    ii += 1
jj = ii
while jj < n and X[jj] == a:
    jj += 1

aa = jj-ii
A = [[a-i for i in X[:ii]], [j-a for j in X[jj:]]]

# print(A)

mm = min(sum(A[0]), sum(A[1]))

AA = [[0] * (mm + 1) for i in range(2)]
for k in range(2):
    AA[k][0] = 1
    for i in range(len(A[k])):
        for j in range(len(AA[k])-1, -1, -1):
            if j - A[k][i] >= 0:
                AA[k][j] += AA[k][j-A[k][i]]
                # print(AA)
                # print(k, j, A[k][i])


# print(AA,aa)
ans = 0
for i in range(1, mm + 1):
    ans += AA[0][i] * AA[1][i]
# print(ans)
ans *= 2 ** aa

print(ans + (2 ** aa - 1))
            
\end{lstlisting} & 
\begin{lstlisting}[style=pythonstyle]
def solve():
    N, A = map(int, input().split())
    x = list(map(int, input().split()))
    
    def count_selections(nums, target_sum, target_count):
        if not nums and target_count == 0 and target_sum == 0:
            return 1
        if not nums or target_count < 0:
            return 0
            
        take = count_selections(nums[1:], target_sum - nums[0], target_count - 1)
        skip = count_selections(nums[1:], target_sum, target_count)
        return take + skip

    ans = 0
    for count in range(1, N + 1):
        if count * A <= sum(x):
            ans += count_selections(x, count * A, count)
    
    print(ans)

solve()
\end{lstlisting} & 
\begin{lstlisting}[style=pythonstyle]
from itertools import combinations

N, A = map(int, input().split())
x = list(map(int, input().split()))

count = 0
for r in range(1, N+1):
    for comb in combinations(x, r):
        if sum(comb) == A * r:
            count += 1
print(count)
\end{lstlisting} \\ \midrule
      \textbf{Qwen} & \textbf{GPT-4o} & \textbf{o3- mini} \\ \hline
\begin{lstlisting}[style=pythonstyle]
from itertools import combinations

def count_ways(N, A, cards):
    total_ways = 0
    for r in range(1, N + 1):
        for combo in combinations(cards, r):
            if sum(combo) == A * r:
                total_ways += 1
    return total_ways

N, A = map(int, input().split())
cards = list(map(int, input().split()))
print(count_ways(N, A, cards))
\end{lstlisting} &
\begin{lstlisting}[style=pythonstyle]
def count_selection_ways(N, A, x):
    dp = [[0] * (sum(x) + 1) for _ in range(N + 1)]
    dp[0][0] = 1
    total_count = 0

    for i in range(N):
        for j in range(i + 1, -1, -1):
            for k in range(sum(x) + 1):
                if dp[j][k] > 0:
                    dp[j + 1][k + x[i]] += dp[j][k]

    for k in range(1, N + 1):
        if k * A <= sum(x):
            total_count += dp[k][k * A]

    return total_count

import sys
input = sys.stdin.read
data = input().split()

N = int(data[0])
A = int(data[1])
x = list(map(int, data[2:]))

print(count_selection_ways(N, A, x))
\end{lstlisting} &
\begin{lstlisting}[style=pythonstyle]
def main():
    import sys
    input_data = sys.stdin.read().split()
    it = iter(input_data)
    N = int(next(it))
    A = int(next(it))
    xs = [int(next(it)) for _ in range(N)]
    max_sum = 2500
    dp = [[0] * (max_sum + 1) for _ in range(N + 1)]
    dp[0][0] = 1
    for x in xs:
        for k in range(N - 1, -1, -1):
            for s in range(max_sum - x, -1, -1):
                dp[k + 1][s + x] += dp[k][s]
    ans = 0
    for k in range(1, N + 1):
        if A * k <= max_sum:
            ans += dp[k][A * k]
    sys.stdout.write(str(ans))

if __name__ == '__main__':
    main()
\end{lstlisting} \\ \midrule
      \textbf{Llama} &   & \\ \hline
\begin{lstlisting}[style=pythonstyle]
from itertools import combinations

def count_ways(N, A, x):
    count = 0
    for r in range(1, N + 1):
        for combo in combinations(x, r):
            if sum(combo) / len(combo) == A:
                count += 1
    return count

N, A = map(int, input().split())
x = list(map(int, input().split()))

print(count_ways(N, A, x))
\end{lstlisting} & & \\ \midrule
     
    \end{tabular}
    \caption{Human written and AI generated Python codes for the average selection problem. We removed comments from DeepSeek V3's code to improve readability.}
    \label{Tab_code}
\end{table*}

\end{document}